\makeatletter
\disable@package@load{program}{}
\makeatother

\documentclass[sn-basic,iicol]{sn-jnl}

\usepackage{graphicx}%
\usepackage{multirow}%
\usepackage{amsmath,amssymb,amsfonts}%
\usepackage{amsthm}%
\usepackage{mathrsfs}%
\usepackage[title]{appendix}%
\usepackage{xcolor}%
\usepackage{textcomp}%
\usepackage{manyfoot}%
\usepackage{booktabs}%
\usepackage{algorithm}%
\usepackage{algorithmicx}%
\usepackage{algpseudocode}%
\usepackage{listings}%
\usepackage{multicol}

\newcommand{\HI}{H\,\textsc{i}}
\newcommand{\HII}{H\,\textsc{ii}}
\newcommand{\CIV}{C\,\textsc{iv}}
\newcommand{\kms}{km s$^{-1}$}
\newcommand{\msun}{\mathrm{M}_\odot}
\newcommand{\msol}{\mathrm{M}_\odot}

\begin{document}

\title[Following the tidal trail]{Following the tidal trail: a history of modeling the Magellanic Stream}


\author*[1,2]{\fnm{Scott} \sur{Lucchini}}\email{scott.lucchini@cfa.harvard.edu}


\affil[1]{\orgname{Center for Astrophysics $\vert$ Harvard \& Smithsonian}, \orgaddress{\street{60 Garden St}, \city{Cambridge}, \state{MA}, \postcode{02438}, \country{USA}}}
\affil[2]{\orgdiv{Department of Physics}, \orgname{University of Wisconsin$-$Madison}, \orgaddress{\city{Madison}, \state{WI}, \postcode{53706}, \country{USA}}}


\abstract{The formation of the Magellanic Stream has puzzled astronomers for decades. In this review, we outline the history of our understanding of the Magellanic System highlighting key observations that have revolutionized thinking of its evolution. We also walk through the major models and theoretical advances that have led to our current paradigm $-$
(1) the LMC and SMC have just had their first pericentric passage around the Milky Way, having approached recently as a bound pair; (2) the LMC and SMC have had several tidal interactions in which material has been stripped out into the Trailing Stream and Leading Arm; (3) the LMC hosted an ionized gas circumgalactic medium which envelops the Clouds and the neutral Stream today, providing the bulk of the associated mass; and (4) the MW's circumgalactic gas provides strong ram pressure and hydrodynamic forces to shape the morphology of the Magellanic System including the formation of a bow shock due to the LMC's supersonic approach.
}


\maketitle

\section{Introduction}\label{intro}

\begin{figure*}
	\centering
	\includegraphics[width=0.8\linewidth]{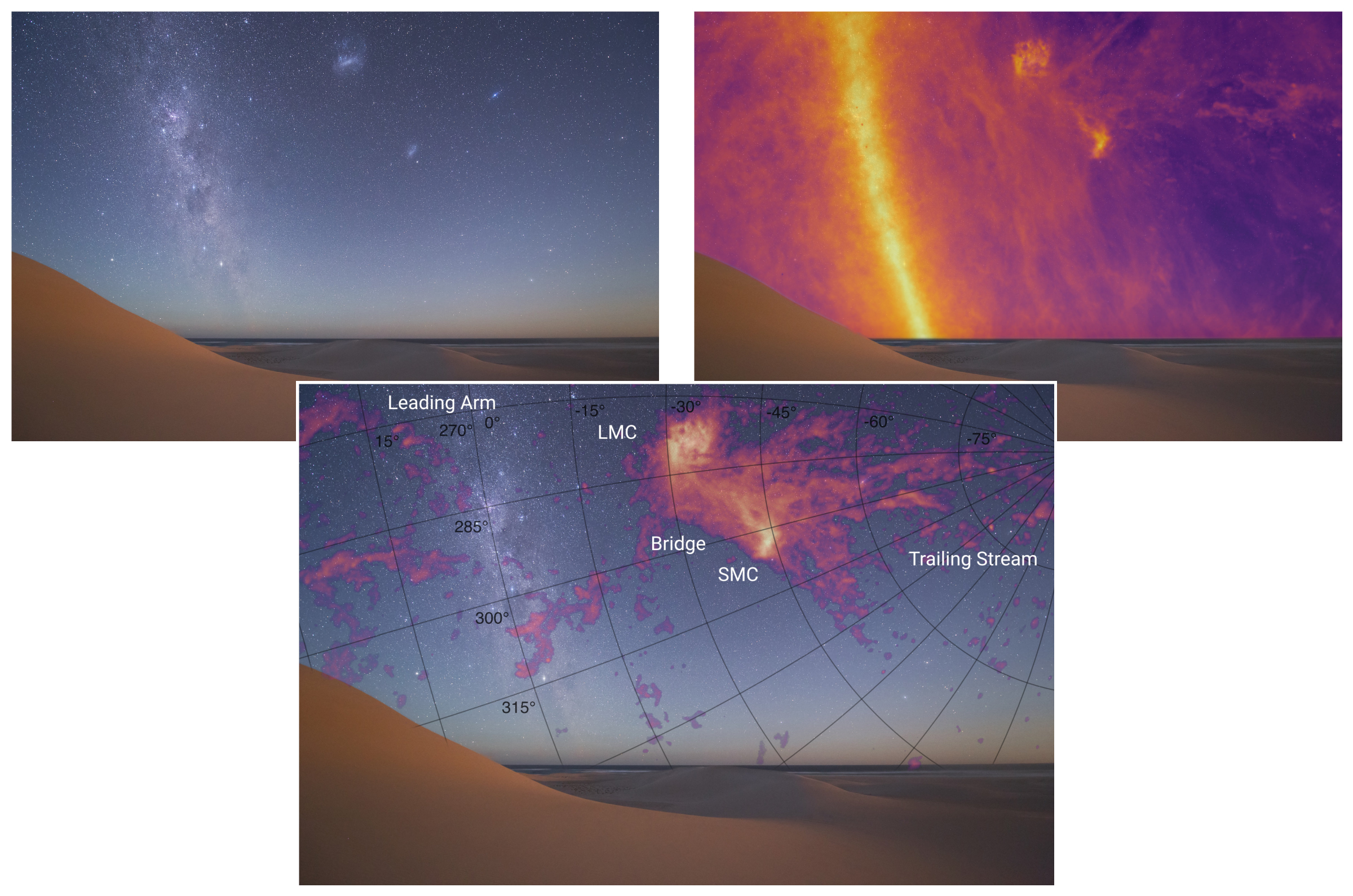}
	\caption{The Magellanic System as it would appear on the sky. The background is a photograph taken by Colin Legg\protect\footnotemark[2]{} in Australia showing the Milky Way disk on the left with the Clouds at the center. The purple and orange overlay shows the 21~cm neutral hydrogen emission from the system. The LMC, Bridge, and SMC are labelled in the center of the image. The Clouds are travelling to the left on the sky and we can see a very extended tail of gas extending behind the Clouds in their orbits, labelled ``Trailing Stream.'' There is also some material out ahead of the LMC and the SMC, labelled ``Leading Arm.'' This image only shows $\sim50\%$ of the length of the Trailing Stream which covers over 100$^\circ$ on the sky in total.}
	\label{fig:sky}
\end{figure*}

The LMC and the SMC (also referred to as the Clouds) are the two most massive dwarf galaxy satellites of the Milky Way (MW). Although faint, the stars in the Clouds are visible to the naked eye from the southern hemisphere and have captivated humans for millennia.
%
They are also the only MW satellites that have managed to retain their gas. The gaseous components of both the Clouds are strongly visible in 21~cm radio emission even on top of the strong Galactic emission from our own gaseous disk. Isolating the high velocity material however, we see so much more. The Magellanic Stream (also referred to as the Stream) is revealed including the Bridge of gas linking the galaxies and the Leading Arm clumps ahead of the Clouds in their orbits. If we could see this neutral \HI\ in our night sky, we would be awestruck by a 200$^\circ$ long network of intertwined filaments dominating the heavens. Figure~\ref{fig:sky} shows a view of the night sky from Australia (top left) overlaid with the total \HI\ emission (top right), and isolating the high velocity material associated with the Clouds (bottom).

Due to the proximity of these stunning galaxies, they have been thoroughly studied academically as well \citep{bok66,donghia16}. The LMC contains several unique features indicative of disequilibrium such as a lopsided bar and a single spiral arm \citep{devaucouleurs72}. It is also host to many well studied star forming regions, the most energetic of which is 30 Doradus which hosts massive, very young star clusters \citep{crowther16,kalari22}. The centroid location of the LMC is difficult to define due to its complicated morphology with different stellar populations providing different locations also inconsistent with the \HI\ center. However, for our purposes, we use the on sky position as defined by the Carbon Stars: (RA, Dec.) = ($80.9^\circ\pm0.3$, $-68.7^\circ\pm0.1$) \citep{wan20}. The distance to the LMC has been constrained to be $49.9\pm2.1$~kpc \citep{degrijs14}. With these values, the 3D Galactocentric position of the LMC can be determined\footnote{The Sun is located at $(x,y,z)=(-8.122,0,0.0208)$~kpc with the Galactic center in the direction of (RA, Dec.) = ($266.4^\circ$, $-28.9^\circ$), and the errors are determined by Monte Carlo sampling. Coordinate transformations have been performed using the astropy library \citep{astropy:2013,astropy:2018,astropy:2022}.} to be $(x,y,z)=(-1.4\pm0.3,-41.4\pm1.6,-27.3\pm1.1)$~kpc. The LMC has a stellar mass of $3.2\times10^9$~$\msol$ \citep{vandermarel09} and a neutral gas mass of $4.4\times10^8$~$\msol$ \citep{bruns05}.

The SMC is a dwarf irregular galaxy with a cigar-shaped structure elongated along the line of sight centered at a distance of $62\pm1.3$~kpc \citep{subramanian12,graczyk20}. The gas and stellar populations exhibit different morphologies, possibly due to the intense interactions between the SMC and the LMC in the past \citep{stanimirovic04,zaritsky00,cioni00}. And as with the LMC, the location of the SMC's center is different for the various stellar populations and the \HI\ center. Using the \HI\ center of (RA, Dec.) = ($15.2^\circ\pm0.4$, $-72.3^\circ\pm0.3$) \citep{diteodoro19}, its 3D Galactocentric position is $(x,y,z)=(15.2\pm0.5,-37.6\pm0.8,-44.1\pm0.9)$. The SMC's stellar and gas masses are $3.1\times10^8$ $\msol$ \citep{stanimirovic04} and $4.0\times10^8$~$\msol$ \citep{bruns05}, respectively.

Also included in the Magellanic System is the 
Bridge, Trailing Stream, and Leading Arm. The Bridge is a gaseous and stellar structure connecting the LMC and the SMC. This structure has historically been considered separately from the Stream because it contains stars while the Stream does not \citep{irwin90}, and the Bridge and the Stream likely formed at different times \citep{besla12,lucchini21}. The stars and the gas in the Bridge have been well studied indicating a SMC-like metallicity \citep{lehner08,misawa09}, a flow of stars from the SMC to the LMC \citep{schmidt20}, and intricate tidal structures such as the Counter-bridge \citep{dias21}. Based on the existence of this Bridge and its proper motions, as well as the observed velocities of the galaxies, the it is very likely that the LMC and SMC experienced a direct collision within the past several hundred million years \citep{zivick18,zivick19,murray19,schmidt20}.

\begin{figure*}
	\centering
	\includegraphics[width=0.7\textwidth]{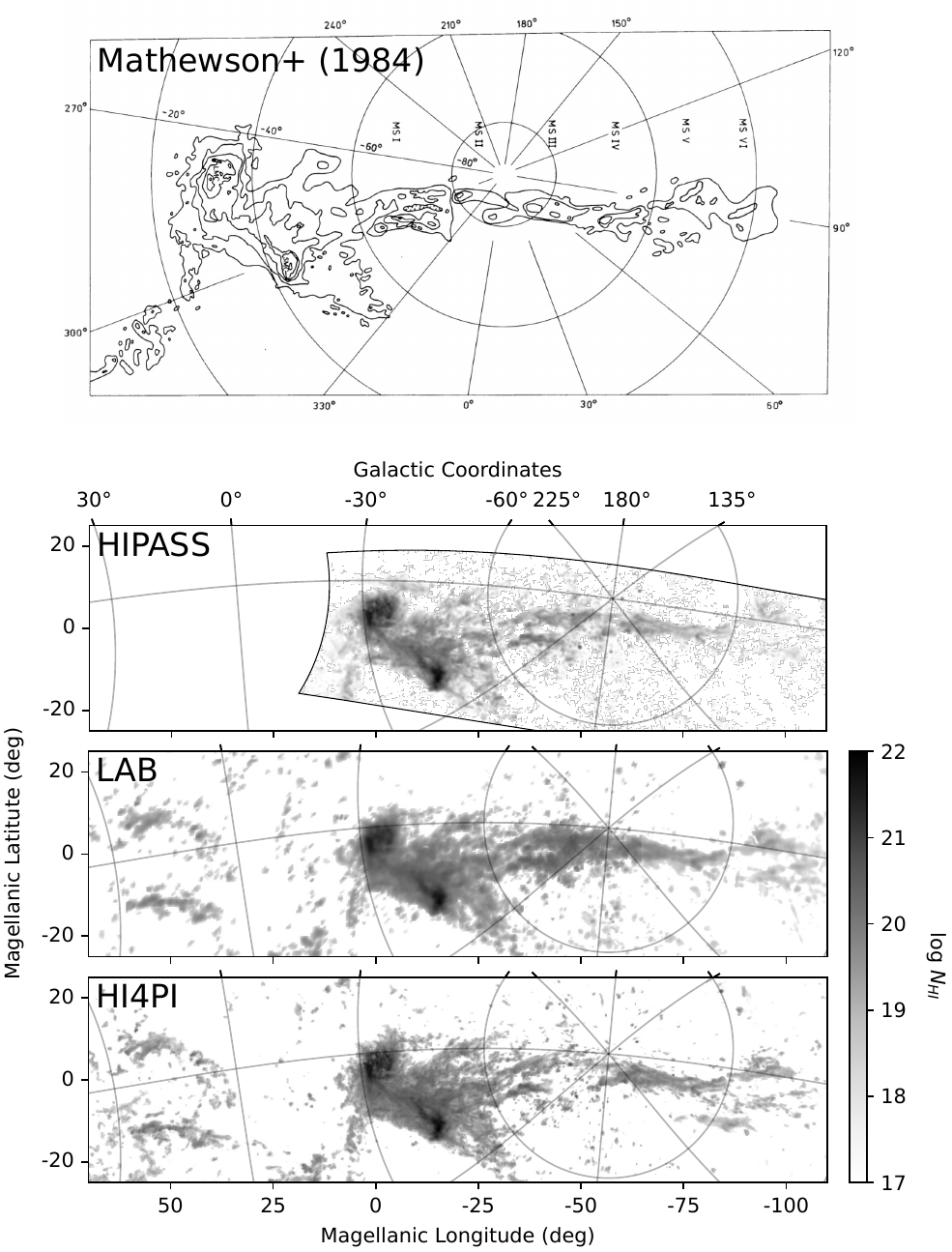}
	\caption{Projections of \HI\ surveys of the LMC/SMC and their surroundings. \textit{Top}: Adapted from \citet{mathewson84} Figure~1 where the knots MS I $-$ VI were first identified. \textit{Center top}: Reprocessed HIPASS data from \citet{putman03}, reprojected into Magellanic Coordinates defined in \citet{nidever08}. \textit{Center bottom}: LAB data showing MC-associated material isolated from Galactic emission using a Gaussian decomposition \citep{nidever08}. \textit{Bottom}: HI4PI high-velocity material \citep{westmeier18}. Note that there is missing low-velocity MC material due to the velocity cuts made. Grid lines on all panels show Galactic coordinates.}
	\label{fig:hi}
\end{figure*}

\footnotetext[2]{\url{https://www.facebook.com/ColinLeggPhotography/}}
The Trailing Stream is the largest extragalactic gaseous structure in our night sky, covering $\sim$100$^\circ$. It is an extended network of interwoven clumpy filaments of gas that originate from within the 
Clouds. Its study has been driven by 21~cm neutral \HI\ observations (\citealt{mathewson74}; Section~\ref{sec:hi}) in combination with absorption line spectroscopy which can reveal its chemistry (\citealt{fox14}; Section~\ref{sec:absorption}). The Leading Arm is the counterpart to the Trailing Stream comprised of clumpy clouds of gas out ahead of the LMC and SMC in their orbits (seen on the left side of Figure~\ref{fig:sky}, bottom). Due to their very high velocities, consistent with the LMC and SMC, it was proposed that these features are tidal material that has been thrown out during the interactions between the Clouds \citep{putman98,besla12,pardy18}.

In this short review we summarize the evolution of our understanding of the formation of the 
Stream. In Section~\ref{sec:obsv}, we highlight some of the key observations that have led to dramatic changes in our picture of the evolution of the Clouds. Section~\ref{sec:models} describes the variety of models that have been proposed over the past 50 years. In Section~\ref{sec:future}, we outline some of the major questions that remain outstanding in this fascinating and complex system. And we conclude in Section~\ref{sec:conclusions}.

\begin{table*}[]
	\centering
	\caption{Observed 3D velocity values for the LMC.}
	\begin{tabular}{cccc}
		\hline
		Reference & $v_r$ (km s$^{-1}$) & $v_t$ (km s$^{-1}$) & $v_\mathrm{tot}$ (km s$^{-1}$) \\\hline
		\citet{kroupa94} & $142\pm220$ & $335\pm220$ & $364\pm138$\\
		\citet{jones94} & $48\pm41$ & $215\pm48$ & $220\pm41$\\
		\citet{kallivayalil06} & $89\pm4$ & $367\pm18$ & $378\pm18$ \\
		\citet{vieira10} & $86\pm48$ & $332\pm52$ & $343\pm48$ \\
		\citet{kallivayalil13} & $64\pm7$ & $314\pm24$ & $321\pm24$ \\\hline
	\end{tabular}
	\label{tab:pms}
\end{table*}
\begin{figure}
	\centering
	\includegraphics[width=1.0\linewidth]{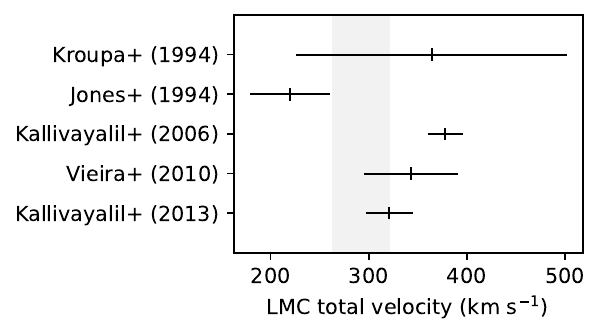}
	\caption{Calculated total 3D velocities for the LMC from observations for several references \citep{kroupa94,jones94,kallivayalil06,vieira10,kallivayalil13}. The grey vertical band denotes the approximate escape velocity at 50~kpc for the MW: $v_\mathrm{esc}=\sqrt{\frac{2GM}{r}}$. For $M=[4,6]\times10^{11}$~$\msol$, we find $v_\mathrm{esc}=[262,321]$~\kms \citep{bland-hawthorn16}.}
	\label{fig:vels}
\end{figure}

\section{Key Observations} \label{sec:obsv}

\subsection{H I} \label{sec:hi}

Over the past fifty years, there have been numerous key observational insights that have led to paradigm shifts in our understanding of the Magellanic System and its evolution. Obviously beginning with the discovery of the Stream and its link to the 
Clouds \citep{wannier72,vankuilenburg72,mathewson74}, observations of the \HI\ have been instrumental in developing models. These maps gave us a picture in broad strokes of the massive Trailing Stream and the six high density ``knots'' along its length \citep[e.g.][]{mathewson84}.

The \HI\ Parkes All Sky Survey (HIPASS; \citealt{barnes01,putman03}) was the first survey of the entire 
Stream at reasonable spatial (15.5') and velocity resolution (26.4 \kms). It had a 3$\sigma$ column density limit of $1.5\times10^{18}$ cm$^{-2}$ per channel. Even though HIPASS was developed to isolate small, discrete structures, \citet{putman03} were able to reprocess the data to include large-scale features in the \HI\ to obtain a full picture of the neutral 
Stream. This work first resolved the Trailing Stream into two distinct filaments that appear intertwined in projection on the sky. Additionally, \citet{putman98} used the HIPASS data to first link the Leading Arm with the 
Clouds due to its high, LMC-like velocity.

\citet{bruns05} conducted a second survey with the Parkes telescope specifically focused on the Stream and Leading Arm. This new data removed the small-scale filtering from HIPASS and was able to obtain much higher velocity resolution of 1.65~\kms\ with a lower column density limit of $4.5\times10^{17}$ cm$^{-2}$ at 3$\sigma$ (with similar spatial resolution of 16'). This survey gave us excellent estimates of the mass across the entire system.

Using the Leiden/Argentine/Bonn (LAB) \HI\ survey \citep{kalberla05}, \citet{nidever08} performed an extensive Gaussian decomposition of the entire Trailing Stream. This survey had a slightly improved velocity resolution of 1.3~\kms, but worse spatial resolution (36') with a similar column density limit of $6.4\times10^{17}$ cm$^{-2}$ at 3$\sigma$. This led to the isolation of the Stream's two filaments kinematically as well as spatially with one leading back to each Cloud.

Finally, the most recent \HI\ data for the Stream comes from the Galactic All-Sky Survey (GASS; \citealt{mcclure-griffiths09}). It is at 16' spatial resolution with 1.0~\kms\ velocity resolution and goes to a 3$\sigma$ depth of $3\times10^{17}$ cm$^{-2}$ per channel. A Gaussian decomposition for this data has yet to be performed, but this survey combines all the best features of the previous works to reveal this stunning nearby system in unprecedented detail. Figure~\ref{fig:hi} shows the HIPASS \citep{putman03}, LAB \citep{nidever08}, and HI4PI (which uses GASS; \citealt{westmeier18,mcclure-griffiths09}) data compared with one of the original maps from \citet{mathewson84}.

\begin{figure*}
	\centering
	\includegraphics[width=0.7\textwidth]{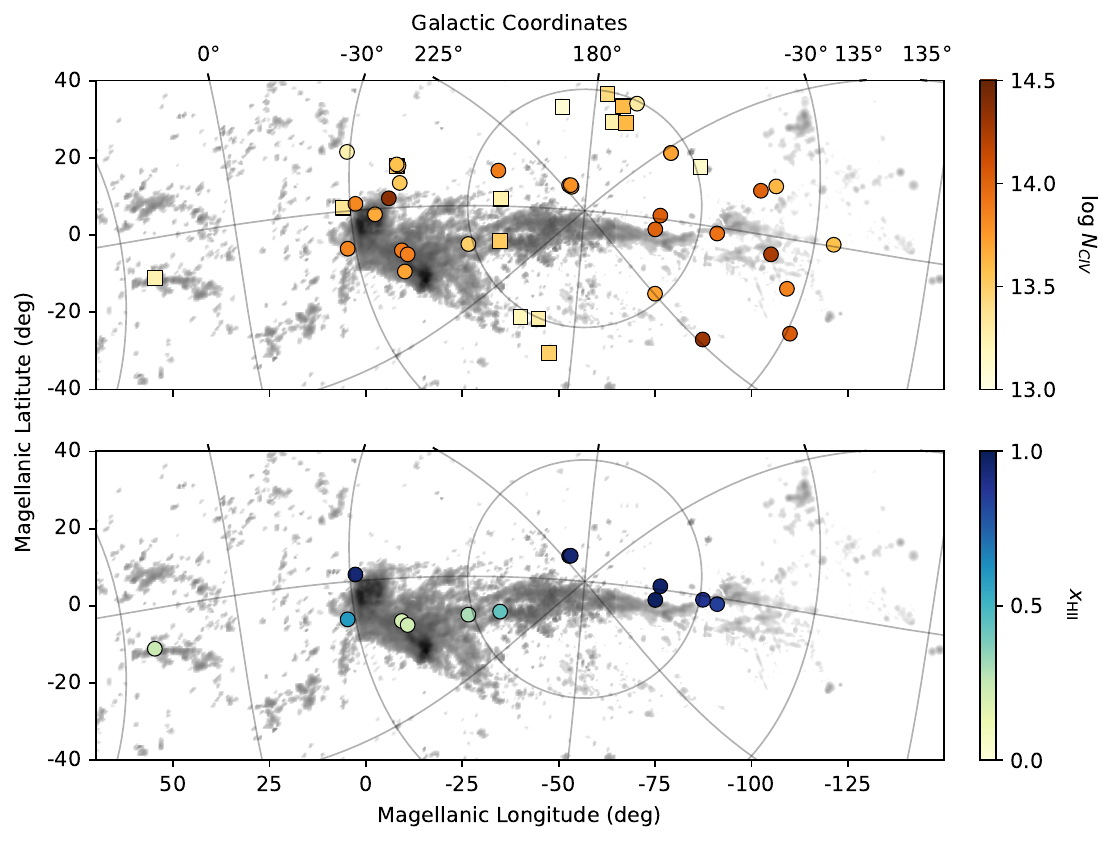}
	\caption{Projection of LAB \HI\ data with properties determined from absorption line spectroscopy overlaid \citep{fox14}. Grid lines on both panels show Galactic coordinates. \textit{Top}: \CIV\ column densities where circles denote detections and squares denote upper limits. \textit{Bottom}: Ionization fraction for the 13 available sight lines, $x_\mathrm{HII}=N(\HII)/\left[N(\HI)+N(\HII)\right]$.}
	\label{fig:civ}
\end{figure*}

\subsection{Proper Motions} \label{sec:pms}

When the Stream was first discovered, the 3D velocities of the Clouds were not known. Precise, long time-baseline observations of stars within the Clouds were required to derive their proper motions. In 1994, two papers came out with estimates from the PPM and the CTIO 4~m \citep{kroupa94,jones94}. However, due to the large error bars on the calculations, the orbital histories of the Clouds were still not well constrained.

The proper motions obtained from the Hubble Space Telescope (HST) were the first to indicate that the LMC may not be bound to our Galaxy \citep{kallivayalil06}. While these velocities were later revised downwards \citep{kallivayalil13}, this work opened the possibility of a ``first-passage'' scenario \citep{besla07}. While the present-day galactic standard of rest velocities of the LMC and SMC are $+84$~\kms\ and $+18$~\kms, respectively (i.e. they are moving away from us), their total velocities are high enough to be unbound to the MW, having just recently had their first pericentric passage of our Galaxy.
A hyperbolic encounter between the LMC and the MW was originally proposed by \citet{lin95}, but it was dismissed at the time since it was assumed that the Clouds were bound to the MW. A summary of a few observations are shown in Table~\ref{tab:pms} and Figure~\ref{fig:vels}.

\citet{besla07} and then \citet{vieira10} performed a comprehensive look at the orbital possibilities of the Clouds and concluded that a first-passage scenario is the most likely not only due to the high tangential velocities, but also taking into account realistic DM profiles for the MW, and because the Clouds are the only nearby gas rich MW satellites. This was a paradigm shift in our understanding of the evolution of the Clouds and the formation of the Stream.

\subsection{UV Absorption Spectroscopy} \label{sec:absorption}

While \HI\ emission has given us a sense of the morphology and velocity of the Stream, UV absorption spectroscopy has illuminated its chemical composition and ionization state. Originally with the \textit{Goddard High Resolution Spectrograph} on HST, metallicities of the Stream gas have constrained the material to have likely originated from within the Clouds, ruling out a primordial origin for the gas \citep{lu94,gibson00}. Furthermore, early studies of other high-velocity gas in the Leading Arm indicated a possible connection to the Clouds \citep{lu94,sembach01} corroborating the results found in \HI\ \citep{putman98}.

Upon the installation of the \textit{Cosmic Origins Spectrograph} on HST, UV spectroscopy 
entered a new era.
\citet{richter13} and \citet{fox13} explored the metallicities of the two Trailing Stream filaments finding chemical distinction on top of the spatial and velocity separation identified in the \HI. \citet{fox14} then used all 69 sight lines to determine a total ionized gas mass associated with the system. They found that, in addition to the $5\times10^8$~$\msun$ of neutral \HI, there was $1.5\times10^9$~$\msun$ of ionized gas with ionization fractions varying from 20\% in the Bridge and on the \HI\ Stream, up to 98\% off the Stream. Figure~\ref{fig:civ} shows the \CIV\ column densities and the ionization fractions. This was again, a paradigm shift as theorists continued to develop models of the formation of the Stream.

\subsection{Total Mass of the LMC} 

Original estimates of the total masses of the Clouds came from their rotation curves \citep{vandermarel02,stanimirovic04,vandermarel14,diteodoro19}. However, these measurements only extend out to 8.7 and 4 kpc for the LMC and the SMC, respectively, while their tidal radii and virial radii are tens of kpc at least. Thus, these must be taken as lower limits for the total masses of these galaxies. These methods give total masses of $1.7\pm0.7 \times10^{10}$ $\msol$ for the LMC \citep{vandermarel14}, and $1.25\pm0.25\times10^9$ $\msol$ for the SMC \citep{diteodoro19}. By assuming a dark matter halo profile for the LMC, its total virial mass can be estimated by extrapolating these observed values out to larger distances. This gives values up to $2.5\times10^{11}$ $\msol$. This is less reliable for the SMC since is has been significantly more distorted through its interactions with the LMC.

Recently, a variety of methods have given insights into the total mass of the LMC indirectly. The six methods outlined below are independent ways to determine the original mass of the LMC (before it approached the MW).

\begin{enumerate}
	
	\item \textit{Abundance Matching}. Through statistics of many observations, the total mass of a galaxy can be assigned based on its stellar mass alone \citep{behroozi10,guo10}. For the LMC's stellar mass of $3.2\times10^9$ $\msol$ \citep{vandermarel09}, we obtain a pre-infall total mass of $\sim1.6\times10^{11}$ $\msol$ \citep{garavito-camargo19}.
	
	\item \textit{Milky Way Reflex Motion}. The fact that the Clouds are their first passage of the MW, combined with the possibility of a high LMC mass, implies that its approach could have significantly distorted our Galaxy. One such implication is that the inner parts of the MW are no longer in alignment with the outer regions \citep{gomez15,petersen20,petersen21}. As the LMC approaches us, the MW will be pulled towards it, leading to an offset between the inner and outer parts of the galaxy.
	Recently, this reflex motion has been detected in observations of the outer stellar halo, requiring that the LMC's total mass be $>10^{11}$ $\msol$ \citep{petersen21}. 
	
	\item \textit{Dark Matter Wake}. In addition to shifting the inner Galaxy with respect to the outer Galaxy, the a first-passage model of the LMC's orbit will induce a more complex ``wake'' in the dark matter distribution of the Milky Way. This effect is comprised of the traditional dynamical friction \citep{chandrasekhar43} combined with a \textit{collective} response visible throughout the Galaxy caused by the resonant responses of the LMC's first passage \citep{garavito-camargo19}. This effect has been investigated in simulations \citep{garavito-camargo19} and detected in observations \citep{conroy21}.
	
	\item \textit{LMC's Satellite Population}. Based on their proper motions, \citet{kallivayalil13} found that the LMC would need to be $>10^{11}$ $\msol$ in order for the two Clouds to remain bound to each other for at least 2 Gyr. In addition to the SMC, the LMC seems to have several other satellites that it is bringing in towards our Galaxy \citep{donghia08,bechtol15,nichols11,pardy20,patel20}. \citet{erkal20} found that, in addition to the SMC, there are six dwarfs likely accreted with the LMC which would require it to have a mass of at least $1.24\times10^{11}$ $\msol$.
	
	\item \textit{Timing Argument}. Due to the expansion of the universe, one can back out the mass of a group of galaxies based on their relative velocities and the fact that their mutual gravitational force has kept them together throughout the age of the universe \citep{lynden-bell81,sandage86,partridge13}. If the LMC is massive, it will necessarily play a role in this evolution and must be considered \citep{gomez15,penarrubia16}. The timing argument indicates an LMC mass of $2.5^{+0.09}_{-0.08}\times10^{11}$ $\msol$ \citep{penarrubia16}. However, the inclusion of the reflex motion of the MW reduces the estimated mass of our Local Group by $\sim10\%$ \citep{chamberlain23}.
	
	\item \textit{Stellar Streams}. The explosion in stellar astrometry from \textit{Gaia} \citep{gaia16} has allowed for unprecedented exploration of our Galaxy's stellar halo. The discovery of numerous stellar streams have given us important insight into the local gravitational potential \citep{ibata94,odenkirchen03,belokurov06}. The deviation from their expected orbits can't be explained by the MW potential alone, so by accounting for the LMC's effects we can constrain its mass to $1-2\times10^{11}$ $\msol$ \citep{erkal19,shipp21}.
	
\end{enumerate}

\begin{figure}[t]
	\centering
	\includegraphics[width=0.5\textwidth]{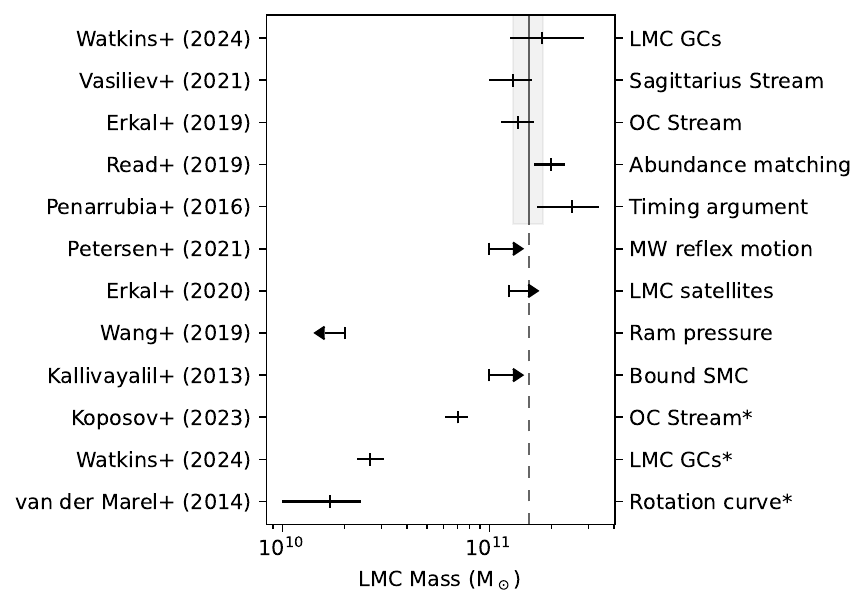}
	\caption{Observational LMC total mass estimates. Literature estimates for the total mass of the LMC. \citet{vandermarel14} and \citet{koposov23} (denoted with asterisks) provide constraints on the mass within 8.7 and 32.4~kpc, respectively, while \citet{watkins24} provides a direct constraint within 13.2~kpc (second from the bottom) in addition to a fit using an NFW profile (top). All other estimates plotted are for the total virial LMC mass. \citet{kallivayalil13}, \citet{erkal20}, and \citet{petersen21} provide lower limits, and \citet{wang19} provide an upper limit. The remaining references \citep{penarrubia16,read19,erkal19,vasiliev21,watkins24} give estimates of the total LMC mass with uncertainties. The error weighted average of these five values is $1.55\pm0.26\times10^{11}$~$\msol$ shown as the vertical line (extended to the bottom of the plot as a dashed line for comparison). The one-sigma errors are shown as a shaded region. The method for determining the LMC mass estimate from each paper is listed on the right edge of the plot.}
    \label{fig:lmcmasses}
\end{figure}

Figure~\ref{fig:lmcmasses} summarizes the different LMC mass estimates from various studies. Three references give estimates within a given radius ($1.7\pm0.7\times10^{10}$ $\msol$ within 8.7~kpc, \citealt{vandermarel14}; $2.7\pm0.4\times10^{10}$ within 13.2~kpc, \citealt{watkins24}; $7.02\pm0.9\times10^{10}$ $\msol$ within 32.8~kpc, \citealt{koposov23}), four references give lower (or upper) limits \citep{kallivayalil13,wang19,erkal20,petersen21}, and five give total mass estimates \citep{penarrubia16,erkal19,shipp21,vasiliev21,watkins24}. Taking the error weighted average of the five total mass measurements, we obtain a value of $1.55\pm0.26\times10^{11}$ $\msol$ shown by the vertical line and shaded region.

\section{Modeling the Formation of the Stream} \label{sec:models}

It is human nature to attempt to explain that which we don't understand. As soon as the Trailing Stream was discovered, explanations of its formation have been developed. In their first paper, \citet{mathewson74} laid out two main formation mechanisms -- tidal debris from gravitational interactions between the Clouds and the MW, or primordial material falling behind the Clouds in their orbit. Further development of these two theories proved fruitful \citep{mathewson76,fujimoto76,lin77} however, major issues remained \citep{bregman79}.

All of these tidal models relied on interactions between the Clouds and the MW over multiple passages. These models had the benefit of easily explaining the disrupted disks of the galaxies as well as the Bridge of gas between them while accurately reproducing their radial velocities. However, this multiple passage tidal model couldn't account for the high negative velocities of the gas observed at the tip of the Stream \citep{lin77}. \citet{gardiner96} and subsequently \citet{yoshizawa03} were able to find an orbital model for the Clouds that could reproduce the global properties of the Stream (including its velocity gradient), Leading Arm, and Bridge through a MW-SMC interaction 1.5 Gyr ago and a LMC-SMC interaction 500 Myr ago. However, these models used analytic potentials for the MW and LMC with a multiple passage scenario.

With the proper motion data of \citet{kallivayalil06}, the first-passage scenario began to be developed (see also Section~\ref{sec:pms}). \citet{besla10,besla12} showed that tidal interactions between the Clouds themselves 
can be sufficient to strip material out into the Trailing Stream, Leading Arm, and Bridge. These simulations were also the first to use live DM potentials for both the LMC and SMC. They also included hydrodynamics and star formation. Additionally, these were the first works to use high masses for the LMC. Previous models used tidally truncated LMC masses (a few $\times10^{10}$~$\msun$), however if the Clouds are on their first passage, the LMC would not have been truncated. They therefore used an LMC with a total mass of $1.8\times10^{11}$~$\msun$ estimated from abundance matching.

Other works also showed that interactions between the Clouds (as opposed to with the MW) can strip material into the Stream even within the context of the multiple-passage scenario \citep{ruzicka10,diaz11}. The orbital history of the Clouds still depended strongly on the total mass of the MW and the shape of its potential. \citet{diaz12} continued developing a second-passage orbit and was able to reproduce bifurcation in the Trailing Stream, along with the Leading Arm with present-day velocities consistent with \citet{vieira10}. Again, this model only followed SMC particles and no hydrodynamics was included.

An alternative model has also been developed based on the ram-pressure scenarios originally proposed by \citet{meurer85} \citep{moore94,mastropietro05}. The modern evolution of this model was presented in \citet{hammer15} in which the LMC and SMC pass through an extended gaseous halo of the MW and undergo a direct collision within the past few hundred Myrs. This model can reproduce the two intertwined filaments in the Trailing Stream and can account for the high ionized gas content detected in absorption \citep{wang19}, however it requires an LMC mass $<2\times10^{10}$~$\msun$. Additionally, the self-consistent formation of the Leading Arm is not possible in a pure ram pressure scenario.

\begin{table*}[]
	{\centering
        \caption{Summary of simulations}
	\begin{tabular}{cccccccc}
		\hline
		Reference & N passes & \multicolumn{3}{c}{Galaxy masses ($10^{10}$~$\msun$)} & Hydro & SF & Cooling \\
		 &  & MW & LMC & SMC &  &  &  \\\hline
		\citet{gardiner96} & 2 & 207$^1$ & 2$^1$ & 0.3 & N & $-$ & $-$ \\
		\citet{yoshizawa03} & 2 & 207$^1$ & 2$^1$ & 0.3 & Y$^2$ & Y & N \\
		\citet{diaz12} & 2 & 173$^1$ & 1$^1$ & 0.3 & N & $-$ & $-$ \\
		\citet{besla12} & 1 & 150$^1$ & 18 & 2.1 & Y & Y & Y \\
            \citet{wang19} & 1 & 50$-$75 & 0.5 & 0.3 & Y & Y & Y \\
            \citet{lucchini21} & 1 & 106 & 19 & 2.1 & Y & Y$^3$ & Y$^3$ \\\hline
	\end{tabular}\\}
        \raggedright{\small\textbf{Notes:}\\
        $^1$ Analytic potentials were used with the MW's position fixed.\\
        $^2$ Used the ``sticky-particle'' method \citep{levinson81}.\\
        $^3$ SF and cooling were included, however stellar and supernova feedback were not so the star formation rates and gas temperatures are not realistic.}
	\label{tab:sims}
\end{table*}

\begin{figure*}
    \centering
    \includegraphics[width=0.8\textwidth]{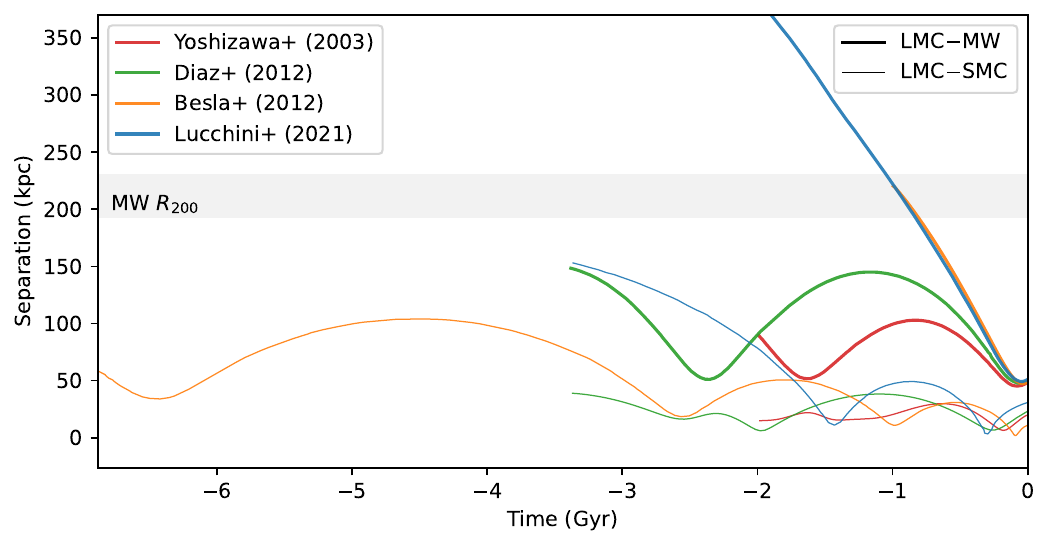}
    \caption{Orbital trajectories for the Clouds for several models. The separation between galaxies is shown as a function of time with present-day at $t=0$. The thick lines are the LMC$-$MW separations, while the thin lines are the LMC$-$SMC separations. The multiple-passage orbits from \citet{yoshizawa03} and \citet{diaz12} are shown in red and green, respectively. The first-passage orbits of \citet{besla12} and \citet{lucchini21} are shown in yellow and blue, respectively. The horizontal grey band denotes the approximate virial radius of the Milky Way ($R_{200}=[192,231]$~kpc; \citealt{bland-hawthorn16}).}
    \label{fig:orbits}
\end{figure*}

\begin{figure*}
    \centering
    \includegraphics[width=0.9\textwidth]{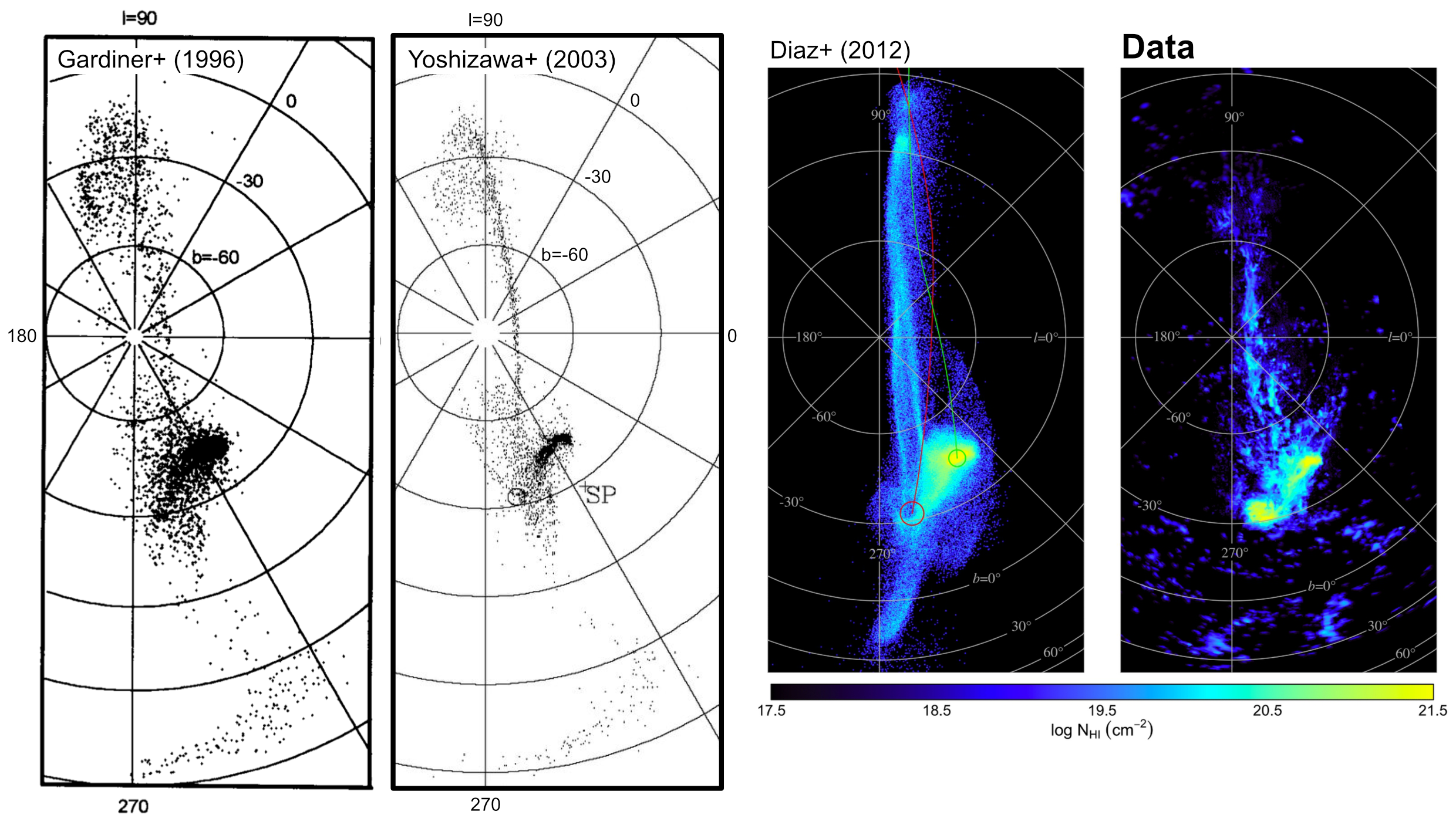}
    \caption{Simulated MC System with second-passage orbits from \citet{gardiner96} (left, adapted from their Figure~6), \citet{yoshizawa03} (left center, adapted from their Figure~8), and \citet{diaz12} (right center, adapted from their Figure~5) compared with the observational data from the LAB survey (right; \citealt{nidever08}).}
    \label{fig:second-passage-models}
\end{figure*}

\begin{figure*}
    \centering
    \includegraphics[width=0.7\textwidth]{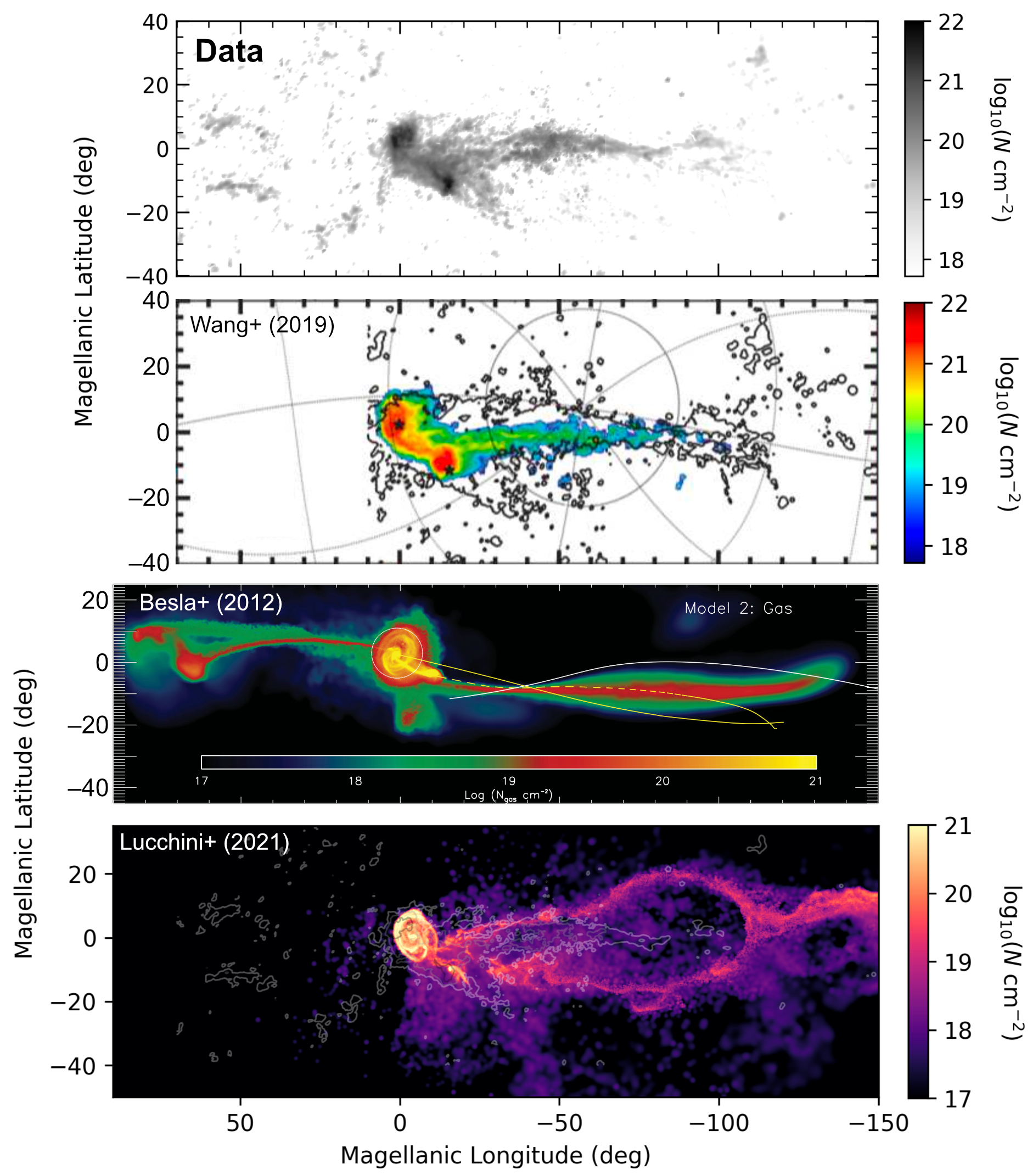}
    \caption{Simulated MC System with first-passage orbits. From top to bottom: LAB data \citep{nidever08}, ram pressure scenario of \citet{wang19} (adapted from their Figure~5), and tidal models of \citet{besla12} (adapted from their Figure~7) and \citet{lucchini21} (adapted from their Figure~3).}
    \label{fig:first-passage-models}
\end{figure*}

\subsection{The Modern Era}

In the past ten years, developments have continued in our understanding of the formation of the 
Stream. However, several studies have focused on specific examples in which our simplified models of the past are insufficient. Approximations made to speed up calculations and reduce computation can have significant effects on the history of the Clouds and the formation of the Stream. Some may argue that we have a picture of the history of the Magellanic System in broad strokes, however as we continue to look more closely at the details, many aspects of our understanding begin to slip away.

First of all, the specific orbits of the Clouds are relatively unconstrained. While we know the positions and velocities of the Clouds quite well, their total masses, as well as the total mass of our own Galaxy, have larger errors. Furthermore, it has been shown that there can be significant differences in the interaction histories of the Clouds when the MW's center of mass is allowed to move \citep{gomez15}. And from observations of the outer stellar halo, we know that our Galaxy is experiencing a reflex motion due to the approach of the LMC \citep{garavito-camargo19,conroy21,petersen21}. Using a fixed MW leads to a preference for first-infall orbits, and recent N-body simulations of the MW and LMC orbital history has shown that there are second-passage models that are consistent with the observed high tangential velocities (and high LMC mass) when taking this reflex motion into consideration \citep{vasiliev24}. Genetic algorithms have been used in an attempt to explore the vast parameter space of orbital histories of the Clouds \citep{guglielmo14}, however including gas physics in these types of studies is difficult due to increased chaos in the interactions as well as increased computational expense. New techniques will be required to explore the possible interaction histories in detail.

Secondly, the inclusion of hydrodynamics is critical. Not only do the gas interactions affect the distribution and velocity of the Stream itself, but the ram pressure interactions between the stripped material and the circumgalactic media (CGM) of the MW and the LMC can play a large role. Moreover, gas physics is necessary for models to be able to account for the ionized component of the Stream which is the dominant component by mass \citep{fox14}. This was first attempted in \citet{pardy18} by increasing the disk masses of the Clouds, however the total mass stripped was still too low. This ionized material seems to be sourced from the LMC's own CGM $-$ the Magellanic Corona \citep{lucchini20,lucchini24,dk22}.

On top of providing the ionized Stream gas, the Corona also leads to hydrodynamical drag on the SMC as it orbits around the LMC, additional heating of the stripped neutral Stream \citep{lucchini21}, and a $\gtrsim$100~kpc long bow shock due to the LMC's supersonic approach to the MW \citep{setton23,carr24}. This shock front could be detectable in column density, X-ray surface brightness, or possibly through spatial variation of the thermal Sunyaev-Zeldovich effect arising from the pressure discontinuity \citep{carr24}. Finally, \citet{tepper-garcia19} found that the formation of the Leading Arm becomes difficult upon the inclusion of the MW's CGM. Follow up studies have shown that creating filamentary structures that survive their passage through the CGM is possible, however large velocities are required and it is unclear if the orbits of the Clouds can provide these strong tidal forces \citep{bustard22}.

\begin{figure*}
	\centering
	\includegraphics[width=1.0\textwidth]{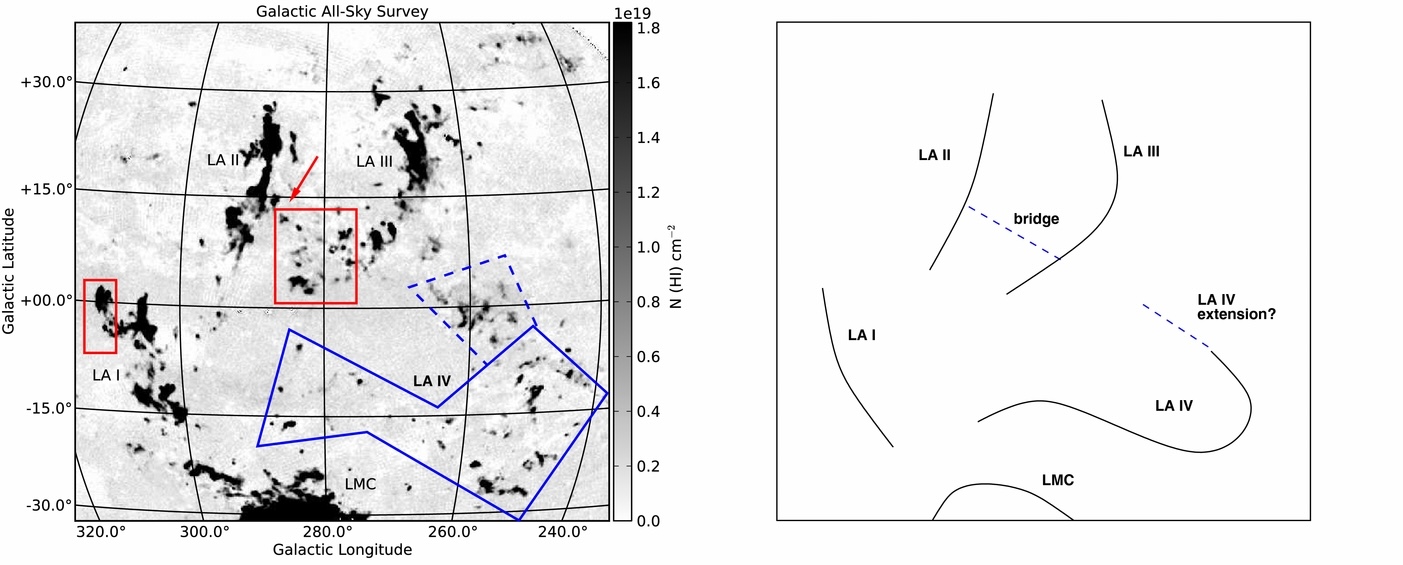}
	\caption{New components of the Leading Arm detected in the GASS survey. \textit{Left}: The integrated \HI\ column density map with the red boxes indicating newly identified extended features of the Leading Arm complexes. The arrow marks a possible bridge between LA II and III, and the blue box outlines a new complex, LA IV, with a possible extension in the dashed blue box. \textit{Right}: A schematic depiction of the Leading Arm features. Adapted from \citet{for13}, Figure~20.}
	\label{fig:la}
\end{figure*}

A summary of these models is presented in Table~\ref{tab:sims} showing the number of LMC passages around the MW, galaxy properties and physics included. Figure~\ref{fig:orbits} also shows the orbital histories of the Clouds for various different models with the LMC-MW separation shown in bold lines and LMC-SMC separation shown as thin lines. The MW's virial radius is shown as the grey band, and this highlights the contrast between second-passage \citep{yoshizawa03,diaz12} and first-passage \citep{besla12,lucchini21} scenarios. Figures~\ref{fig:second-passage-models} and \ref{fig:first-passage-models} show the appearance of the Stream for the different models discussed above. The results of the second-passage models from \citet{gardiner96}, \citet{yoshizawa03}, and \citet{diaz12} are shown in Figure~\ref{fig:second-passage-models} in comparison to the \HI\ LAB data \citep{nidever08}. These models can reproduce the length and bifurcation of the Trailing Stream as well as the Leading Arm. Figure~\ref{fig:first-passage-models} shows the appearance of the Stream for the first-passage models. The ram-pressure scenario of \citet{wang19} reproduces the morphology of the Stream quite well while also forming sufficient ionized material due to heating via Kelvin-Helmholtz instabilities. \citet{besla12} self-consistently forms a Leading Arm while reproducing features and kinematics of the LMC disk. Finally, \citet{lucchini21}, including circumgalactic gas around the LMC and MW, is also able to reproduce the ionized gas mass as well as the turbulent, filamentary morphology with a high-mass LMC.

\begin{figure*}
	\centering
	\includegraphics[width=0.8\textwidth]{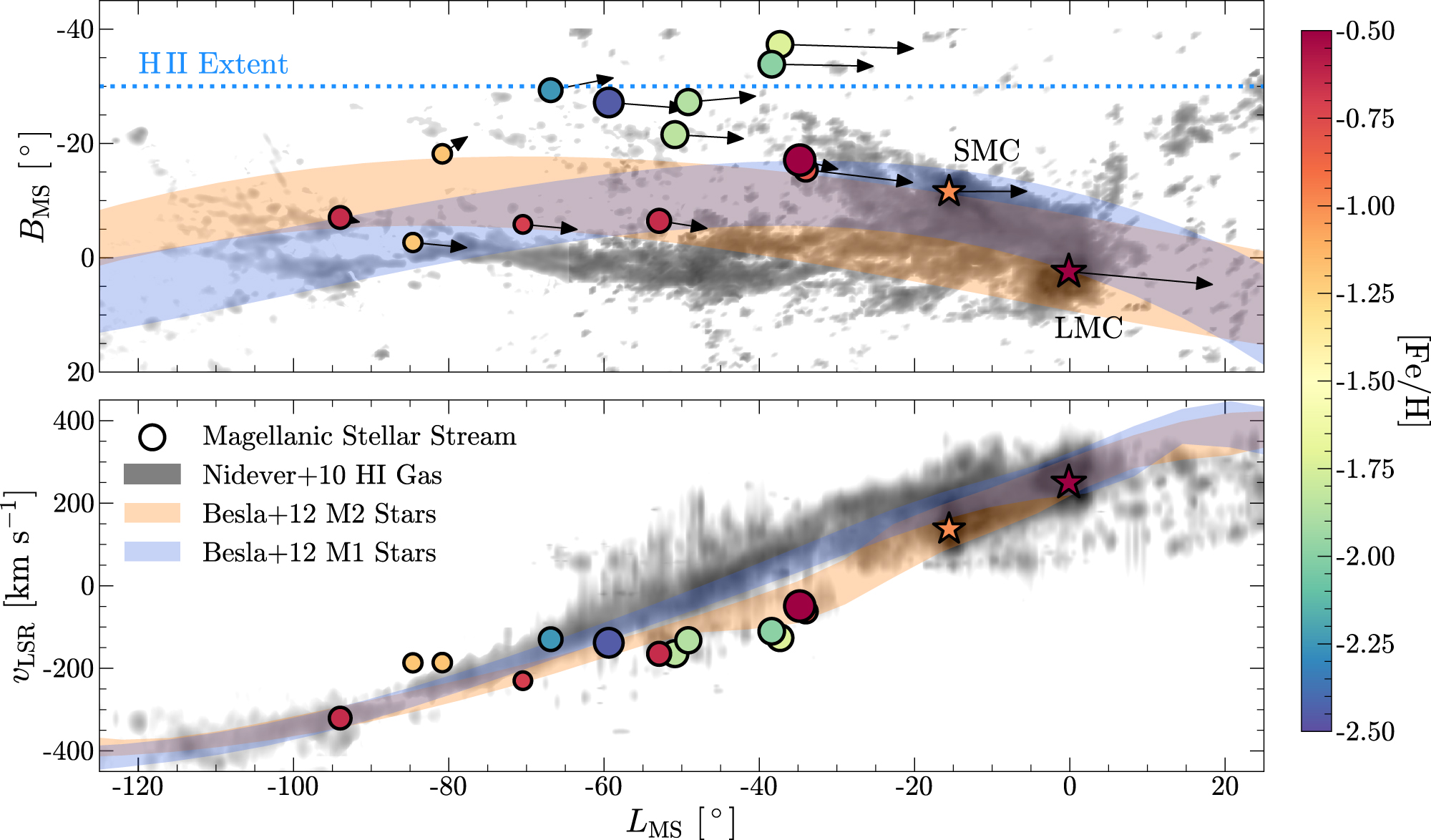}
	\caption{A comparison of the Magellanic stellar stream from \citet{chandra23} with the \HI\ data from \citet{nidever10}. The gas is shown in grey with the detected stars plotted with circles colored by metallicity. \textit{Top}: Locations on the sky (in inverted Magellanic Coordinates) with arrows also denote their proper motions. \textit{Bottom}: LSR velocity vs Longitude. In both panels, the colored regions indicate predictions from the simulations in \citet{besla12}. Adapted from \citet{chandra23}, Figure~4.}
	\label{fig:stellar}
\end{figure*}

\section{Looking Ahead} \label{sec:future}

Over fifty years of observing and modeling the Magellanic System we have come to understand a significant amount about its origins and evolution. The LMC's high 3D velocity implies that the Clouds are on their first passage around our Galaxy \citep{besla07}. Indications that the LMC had a mass $>10^{11}$ $\msun$ combined with the first-infall trajectory tell us that the neutral Trailing Stream must have been stripped via tidal interactions between the Clouds \citep{besla12}. Furthermore, the massive amount of ionized material associated with the Clouds and Stream seems to have originated in the LMC's circumgalactic medium, the Magellanic Corona \citep{lucchini20}.

However, there are still many outstanding questions:
\begin{enumerate}
    \item What is the true nature and origin of the Leading Arm?
    \item How far away is the gas in the Trailing Stream?
    \item What is the ionization mechanism for the Stream?
    \item What can we learn from the stellar component of the Stream?
    \item What is the precise interaction history between the Clouds?
    \item Can we find evidence of the LMC's bow shock?
\end{enumerate}

\noindent\textbf{Leading Arm:}\\
The Leading Arm was originally identified as a component of the Magellanic System based on its velocities ahead of the Clouds in their orbits \citep{putman98}. It is comprised of three substructures (LA I, II, and III; \citealt{putman98,bruns05,for13}) with LA I physically and kinematically connected to the LMC \citep{nidever10} and LA III as the most distant. We do have absorption spectroscopy data for a few sight lines in the Leading Arm, but the metallicities are not well constrained \citep{lu98,fox18}. The O/H values are low for all the leading substructures with LMC origin ruled out for LA III \citep{fox18}. Distances to the clouds have been estimated at $\sim$20~kpc based on detected stellar populations \citep{casetti-dinescu14}, apparent interactions with the MW disk \cite{mcclure-griffiths08}, and H$\alpha$ emission \citep{antwi-danso20}.

Originally thought to be a smoking gun detection confirming a tidal origin of the Stream, further exploration has shown that the existence of the Leading Arm is complicated even in tidal models \citep{tepper-garcia19}. The intense ram pressure experienced by the leading gas passing through the MW CGM can easily break up the clouds or push them back into the trailing region \citep{bustard22}. Moreover, it has been proposed that the leading clouds could be stripped gas from other dwarfs that were leading the Clouds in their infall towards the MW \citep{yang14,hammer15}. This would be consistent with a ram-pressure formation of the Trailing Stream, however viable candidates for the remnants of these stripped dwarfs have yet to be identified \citep{tepper-garcia19}.

Moving forward, there are two questions that remain unanswered: (i) did the Leading Arm originate from the Clouds? and (ii) how do the Leading Arm substructures survive? A closer look at the latest observations (in \HI\ and absorption spectra) will help illuminate the constraints for the Leading Arm's origin. Recent work with the GASS survey \citep{for13} has identified new possible connections between the Leading clumps increasing evidence their origins tracing back to the Clouds. Figure~\ref{fig:la} shows some of these features. And a broader parameter space exploration of orbital histories of the Clouds will allow us to determine if the tidally stripped material can actually survive to the present day.\\

\begin{figure*}
	\centering
	\includegraphics[width=1.0\textwidth]{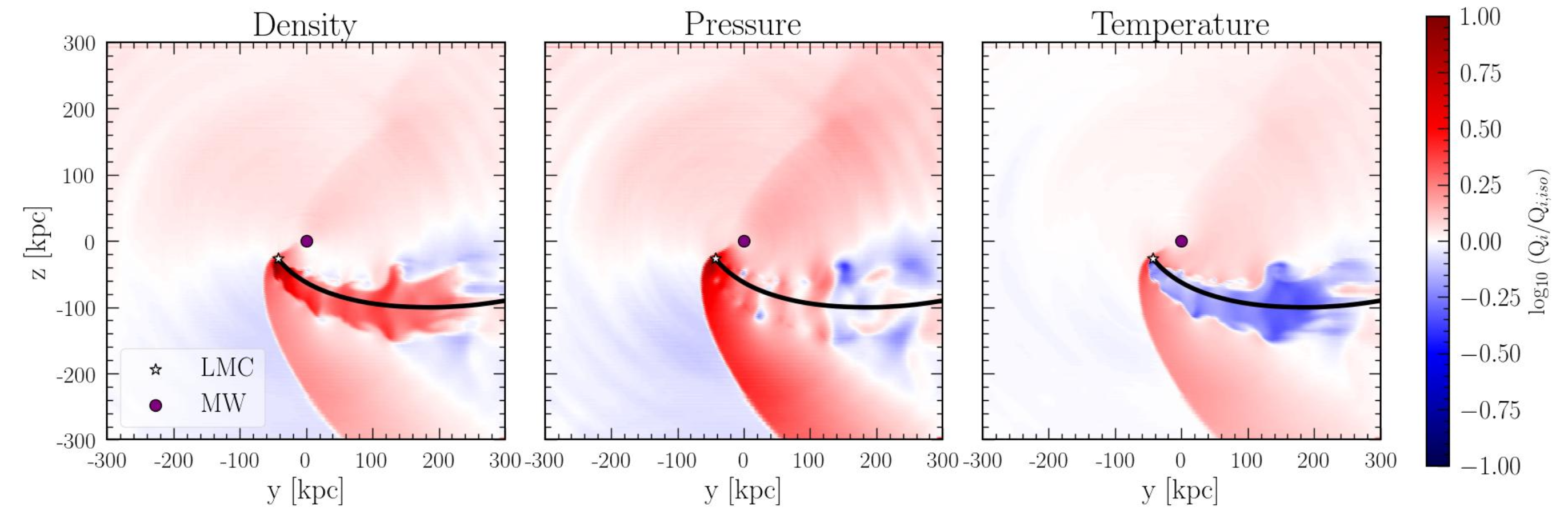}
	\caption{Changes in properties of the MW CGM due to the infall of the LMC including its own CGM (the Magellanic Corona). Density (\textit{left}), pressure (\textit{center}), and temperature (\textit{right}) changes with respect to the isolated MW. All panels are shown in the Galactocentric y$-$z plane in which the MW disk is edge-on. The LMC's present-day position is marked as a white star and the MW is marked with a black circle. The bow shock is clearly visible with other features such as the wake and collective response appearing as well. Adapted from \citet{carr24}, Figure~4.}
	\label{fig:shock}
\end{figure*}

\noindent\textbf{Distance to the gas:}\\
We have no way to directly measure distances to gas in astronomy. Modern techniques such as 3D dust mapping can give us clues about the gas locations, however these datasets only extend to a few kpc from the Sun \citep{dharmawardena24,edenhofer24}. Indirect estimates can come from H$\alpha$ luminosities, however this relies on radiation models of our Galaxy (and the Clouds; \citealt{barger17,antwi-danso20}. We believe the most promising path forward here will come through distance bracketed absorption spectroscopy using stars in the MW halo \citep[e.g.][]{lehner22}. Currently, this technique is limited since it requires bright halo stars with strong distance estimates closely spaced on the sky. However, with next-generation observatories, magnitude limits, velocity resolution, and number of sources will only continue to increase, making this method viable for mapping the gas in and around the MW.\\

\noindent\textbf{Ionization Mechanism:}\\
UV absorption observations have shown that there is ionized gas in and around the Stream, however what ionized it remains an open question. Depending on the ionization mechanism, the mass estimates of the ionized material could change. This ionization could come from many different sources, with each probably contributing a component. One theory arises from evidence of a recent Seyfert flare from the MW \citep{bland-hawthorn13,bland-hawthorn19}. H$\alpha$ observations along the Stream indicate increased intensities below the southern Galactic pole, which could have been hit with strong ionizing radiation from a recent Galactic flare also correlated with the Fermi bubbles \citep{bland-hawthorn03,su10}. Another proposition involves ionization from a shock cascade as the Stream moves through ambient gas in the LMC or MW CGMs \citep{bland-hawthorn07}. Finally, ionization from the MW (and LMC) stellar radiation field could also heat the low-density Stream material \citep{barger17}. However, the effectiveness of this mechanism depends strongly on the distance to the Stream \citep{lucchini21}. More detailed observations of the variability in and structure of the ionized Stream material will help disentangle the different processes at work. This will ultimately give us a more complete picture of the total amount of material that has fallen in with the LMC and SMC and how our Galaxy will be affected in the future.\\

\noindent\textbf{Stellar stream:}\\
An additional hint about the distance to the gas could come from information about a stellar stream linked to the Clouds \citep{belokurov16,zaritsky20,chandra23}. The most recent detection of such a stream in \citet{chandra23} finds stars at 60$-$120~kpc with kinematics and chemistry matching the Clouds (Figure~\ref{fig:stellar}). While the physics affecting the collisionless stars will be different from the gas, these observations can give us indications of where the gaseous Trailing Stream could be in 3D space. Moreover, the existence and properties of these stars themselves provides strong constraints on models. However, with only 13 stars identified so far, more work will be required to connect the observations to the models.\\

\noindent\textbf{Interaction history:}\\
Directly integrating the past orbital history of the MW, LMC, and SMC is not yet constraining due to the errors in the Cloud velocities and masses of the galaxies. Additional pieces of information are required to determine the interactions between the Clouds. Many recent studies have identified complex features in the outskirts of the Clouds that are most likely indications of their past encounters \citep{dias22,choi22,cheng22,elyoussoufi23,navarrete23}. The existence and properties of the gaseous Bridge also indicate a very recent ($\lesssim$200~Myr ago) close (5$-$10~kpc impact parameter) passage between the Clouds \citep{choi22,oliveira23}.

Going back further than $\sim$1~Gyr requires looking at structures on larger scales. Access to extremely high resolution \HI\ maps of the Trailing Stream and Leading Arm can give us hints about the Clouds' interactions as well. Unfortunately, due to the complexities of the system, retrieving orbital histories from these features requires forward modeling of our simulations. Therefore one of the major hurdles that we face in coming years is determining an efficient method of parameter space exploration.\\

\noindent\textbf{Bow shock:}\\
Originally purported in \citet{setton23}, the LMC's high velocity, recent approach to the MW should induce a shock as it passes through the MW CGM. \citet{carr24} also showed that the extent of this bow shock would only be increased upon inclusion of the LMC's Corona (Figure~\ref{fig:shock}). Observational evidence of this shock could provide crucial information about the properties of the LMC and MW. Specifically, the standoff radius of the shock would give a direct estimate of the ratio of the MW's CGM mass to the LMC's CGM mass. Future X-ray emission observatories could also directly observe the increase in high temperature material in this shock.

\section{Conclusions} \label{sec:conclusions}

The Magellanic System has been an incredibly fruitful laboratory to study galaxy dynamics, galaxy evolution, gas physics, accretion, and metal transport over the past fifty years. And yet, we still have much to glean from this serendipitously nearby system. Future observations will unveil the details of the diffuse ionized component of the Stream, the distance to the gas, and the small-scale, low column density features within the Clouds and Stream. Next generation simulations and models will require new techniques to efficiently traverse the vast parameter space of orbital histories for the Clouds. However, upon success, we will continue to gain understanding of the reasons these galaxies appear so unique with complex features and properties. N-body models will disentangle the tidal features within the disks and help us understand the true nature of an associated stellar stream. Hydrodynamic models will illuminate the full impact of the LMC Corona on the evolution of the Clouds as well as on our Galaxy.

\onecolumn{
\begin{multicols}{2}
\backmatter

\bmhead{Acknowledgements}

SL acknowledges the support of his PhD thesis advisor, Elena D'Onghia, and his close collaborator, Andy Fox, throughout his graduate work studying the Magellanic Stream, and would also like to thank them both for their constructive comments on the manuscript. SL thanks Mary Putman and David Nidever for providing the data from the HIPASS and LAB surveys for reproduction in this work. SL would also like to thank Kenji Bekki, Gurtina Besla, Chris Carr, Vedant Chandra, Bi-Qing For, Francois Hammer, and Masafumi Noguchi, for allowing the use of their figures in this review.




\end{multicols}

\bibliography{references}

\def\apj{ApJ} \def\apjs{ApJ Supplement} \def\apjl{ApJ Letters}
  \def\mnras{MNRAS} \def\araa{ARA\&A} \def\aap{A\&A} \def\aj{AJ}
  \def\textdegree{$^\circ$} \def\nat{Nature} \def\annrev{ARA\&A}
  \def\pasa{PASA} \def\pasj{PASJ}
\begin{thebibliography}{136}
\providecommand{\natexlab}[1]{#1}
\providecommand{\url}[1]{{#1}}
\providecommand{\urlprefix}{URL }
\providecommand{\doi}[1]{\url{https://doi.org/#1}}
\providecommand{\eprint}[2][]{\url{#2}}
 \bibcommenthead

\bibitem[{{Antwi-Danso} et~al(2020){Antwi-Danso}, {Barger}, and
  {Haffner}}]{antwi-danso20}
{Antwi-Danso} J, {Barger} KA, {Haffner} LM (2020) {H{\ensuremath{\alpha}}
  Distances to the Leading Arm of the Magellanic Stream}. \apj 891(2):176.
  \doi{10.3847/1538-4357/ab6ef9},
  {\href{https://arxiv.org/abs/2004.09606}{{arXiv:2004.09606}}} {[astro-ph.GA]}

\bibitem[{{Astropy Collaboration} et~al(2013){Astropy Collaboration},
  {Robitaille}, {Tollerud}, {Greenfield}, {Droettboom}, {Bray}, {Aldcroft},
  {Davis}, {Ginsburg}, {Price-Whelan}, {Kerzendorf}, {Conley}, {Crighton},
  {Barbary}, {Muna}, {Ferguson}, {Grollier}, {Parikh}, {Nair}, {Unther},
  {Deil}, {Woillez}, {Conseil}, {Kramer}, {Turner}, {Singer}, {Fox}, {Weaver},
  {Zabalza}, {Edwards}, {Azalee Bostroem}, {Burke}, {Casey}, {Crawford},
  {Dencheva}, {Ely}, {Jenness}, {Labrie}, {Lim}, {Pierfederici}, {Pontzen},
  {Ptak}, {Refsdal}, {Servillat}, and {Streicher}}]{astropy:2013}
{Astropy Collaboration}, {Robitaille} TP, {Tollerud} EJ, et~al (2013) {Astropy:
  A community Python package for astronomy}. \aap 558:A33.
  \doi{10.1051/0004-6361/201322068},
  {\href{https://arxiv.org/abs/1307.6212}{{arXiv:1307.6212}}} {[astro-ph.IM]}

\bibitem[{{Astropy Collaboration} et~al(2018){Astropy Collaboration},
  {Price-Whelan}, {Sip{\H{o}}cz}, {G{\"u}nther}, {Lim}, {Crawford}, {Conseil},
  {Shupe}, {Craig}, {Dencheva}, {Ginsburg}, {Vand erPlas}, {Bradley},
  {P{\'e}rez-Su{\'a}rez}, {de Val-Borro}, {Aldcroft}, {Cruz}, {Robitaille},
  {Tollerud}, {Ardelean}, {Babej}, {Bach}, {Bachetti}, {Bakanov}, {Bamford},
  {Barentsen}, {Barmby}, {Baumbach}, {Berry}, {Biscani}, {Boquien}, {Bostroem},
  {Bouma}, {Brammer}, {Bray}, {Breytenbach}, {Buddelmeijer}, {Burke},
  {Calderone}, {Cano Rodr{\'\i}guez}, {Cara}, {Cardoso}, {Cheedella}, {Copin},
  {Corrales}, {Crichton}, {D'Avella}, {Deil}, {Depagne}, {Dietrich}, {Donath},
  {Droettboom}, {Earl}, {Erben}, {Fabbro}, {Ferreira}, {Finethy}, {Fox},
  {Garrison}, {Gibbons}, {Goldstein}, {Gommers}, {Greco}, {Greenfield},
  {Groener}, {Grollier}, {Hagen}, {Hirst}, {Homeier}, {Horton}, {Hosseinzadeh},
  {Hu}, {Hunkeler}, {Ivezi{\'c}}, {Jain}, {Jenness}, {Kanarek}, {Kendrew},
  {Kern}, {Kerzendorf}, {Khvalko}, {King}, {Kirkby}, {Kulkarni}, {Kumar},
  {Lee}, {Lenz}, {Littlefair}, {Ma}, {Macleod}, {Mastropietro}, {McCully},
  {Montagnac}, {Morris}, {Mueller}, {Mumford}, {Muna}, {Murphy}, {Nelson},
  {Nguyen}, {Ninan}, {N{\"o}the}, {Ogaz}, {Oh}, {Parejko}, {Parley}, {Pascual},
  {Patil}, {Patil}, {Plunkett}, {Prochaska}, {Rastogi}, {Reddy Janga},
  {Sabater}, {Sakurikar}, {Seifert}, {Sherbert}, {Sherwood-Taylor}, {Shih},
  {Sick}, {Silbiger}, {Singanamalla}, {Singer}, {Sladen}, {Sooley},
  {Sornarajah}, {Streicher}, {Teuben}, {Thomas}, {Tremblay}, {Turner},
  {Terr{\'o}n}, {van Kerkwijk}, {de la Vega}, {Watkins}, {Weaver}, {Whitmore},
  {Woillez}, {Zabalza}, and {Astropy Contributors}}]{astropy:2018}
{Astropy Collaboration}, {Price-Whelan} AM, {Sip{\H{o}}cz} BM, et~al (2018)
  {The Astropy Project: Building an Open-science Project and Status of the v2.0
  Core Package}. \aj 156(3):123. \doi{10.3847/1538-3881/aabc4f},
  {\href{https://arxiv.org/abs/1801.02634}{{arXiv:1801.02634}}} {[astro-ph.IM]}

\bibitem[{{Astropy Collaboration} et~al(2022){Astropy Collaboration},
  {Price-Whelan}, {Lim}, {Earl}, {Starkman}, {Bradley}, {Shupe}, {Patil},
  {Corrales}, {Brasseur}, {N{"o}the}, {Donath}, {Tollerud}, {Morris},
  {Ginsburg}, {Vaher}, {Weaver}, {Tocknell}, {Jamieson}, {van Kerkwijk},
  {Robitaille}, {Merry}, {Bachetti}, {G{"u}nther}, {Aldcroft},
  {Alvarado-Montes}, {Archibald}, {B{'o}di}, {Bapat}, {Barentsen}, {Baz{'a}n},
  {Biswas}, {Boquien}, {Burke}, {Cara}, {Cara}, {Conroy}, {Conseil}, {Craig},
  {Cross}, {Cruz}, {D'Eugenio}, {Dencheva}, {Devillepoix}, {Dietrich},
  {Eigenbrot}, {Erben}, {Ferreira}, {Foreman-Mackey}, {Fox}, {Freij}, {Garg},
  {Geda}, {Glattly}, {Gondhalekar}, {Gordon}, {Grant}, {Greenfield}, {Groener},
  {Guest}, {Gurovich}, {Handberg}, {Hart}, {Hatfield-Dodds}, {Homeier},
  {Hosseinzadeh}, {Jenness}, {Jones}, {Joseph}, {Kalmbach}, {Karamehmetoglu},
  {Ka{l}uszy{'n}ski}, {Kelley}, {Kern}, {Kerzendorf}, {Koch}, {Kulumani},
  {Lee}, {Ly}, {Ma}, {MacBride}, {Maljaars}, {Muna}, {Murphy}, {Norman},
  {O'Steen}, {Oman}, {Pacifici}, {Pascual}, {Pascual-Granado}, {Patil},
  {Perren}, {Pickering}, {Rastogi}, {Roulston}, {Ryan}, {Rykoff}, {Sabater},
  {Sakurikar}, {Salgado}, {Sanghi}, {Saunders}, {Savchenko}, {Schwardt},
  {Seifert-Eckert}, {Shih}, {Jain}, {Shukla}, {Sick}, {Simpson},
  {Singanamalla}, {Singer}, {Singhal}, {Sinha}, {Sip{H{o}}cz}, {Spitler},
  {Stansby}, {Streicher}, {{{S}}umak}, {Swinbank}, {Taranu}, {Tewary},
  {Tremblay}, {Val-Borro}, {Van Kooten}, {Vasovi{'c}}, {Verma}, {de Miranda
  Cardoso}, {Williams}, {Wilson}, {Winkel}, {Wood-Vasey}, {Xue}, {Yoachim},
  {Zhang}, {Zonca}, and {Astropy Project Contributors}}]{astropy:2022}
{Astropy Collaboration}, {Price-Whelan} AM, {Lim} PL, et~al (2022) {The Astropy
  Project: Sustaining and Growing a Community-oriented Open-source Project and
  the Latest Major Release (v5.0) of the Core Package}. \apj 935(2):167.
  \doi{10.3847/1538-4357/ac7c74},
  {\href{https://arxiv.org/abs/2206.14220}{{arXiv:2206.14220}}} {[astro-ph.IM]}

\bibitem[{{Barger} et~al(2017){Barger}, {Madsen}, {Fox}, {Wakker},
  {Bland-Hawthorn}, {Nidever}, {Haffner}, {Antwi-Danso}, {Hernand ez},
  {Lehner}, {Hill}, {Curzons}, and {Tepper-Garc{\'\i}a}}]{barger17}
{Barger} KA, {Madsen} GJ, {Fox} AJ, et~al (2017) {Revealing the Ionization
  Properties of the Magellanic Stream Using Optical Emission}. \apj 851(2):110.
  \doi{10.3847/1538-4357/aa992a},
  {\href{https://arxiv.org/abs/1711.04395}{{arXiv:1711.04395}}} {[astro-ph.GA]}

\bibitem[{{Barnes} et~al(2001){Barnes}, {Staveley-Smith}, {de Blok},
  {Oosterloo}, {Stewart}, {Wright}, {Banks}, {Bhathal}, {Boyce}, {Calabretta},
  {Disney}, {Drinkwater}, {Ekers}, {Freeman}, {Gibson}, {Green}, {Haynes}, {te
  Lintel Hekkert}, {Henning}, {Jerjen}, {Juraszek}, {Kesteven}, {Kilborn},
  {Knezek}, {Koribalski}, {Kraan-Korteweg}, {Malin}, {Marquarding}, {Minchin},
  {Mould}, {Price}, {Putman}, {Ryder}, {Sadler}, {Schr{\"o}der}, {Stootman},
  {Webster}, {Wilson}, and {Ye}}]{barnes01}
{Barnes} DG, {Staveley-Smith} L, {de Blok} WJG, et~al (2001) {The HI Parkes All
  Sky Survey: southern observations, calibration and robust imaging}. \mnras
  322(3):486--498. \doi{10.1046/j.1365-8711.2001.04102.x}

\bibitem[{{Bechtol} et~al(2015){Bechtol}, {Drlica-Wagner}, {Balbinot},
  {Pieres}, {Simon}, {Yanny}, {Santiago}, {Wechsler}, {Frieman}, {Walker},
  {Williams}, {Rozo}, {Rykoff}, {Queiroz}, {Luque}, {Benoit-L{\'e}vy},
  {Tucker}, {Sevilla}, {Gruendl}, {da Costa}, {Fausti Neto}, {Maia}, {Abbott},
  {Allam}, {Armstrong}, {Bauer}, {Bernstein}, {Bernstein}, {Bertin}, {Brooks},
  {Buckley-Geer}, {Burke}, {Carnero Rosell}, {Castander}, {Covarrubias},
  {D'Andrea}, {DePoy}, {Desai}, {Diehl}, {Eifler}, {Estrada}, {Evrard},
  {Fernandez}, {Finley}, {Flaugher}, {Gaztanaga}, {Gerdes}, {Girardi},
  {Gladders}, {Gruen}, {Gutierrez}, {Hao}, {Honscheid}, {Jain}, {James},
  {Kent}, {Kron}, {Kuehn}, {Kuropatkin}, {Lahav}, {Li}, {Lin}, {Makler},
  {March}, {Marshall}, {Martini}, {Merritt}, {Miller}, {Miquel}, {Mohr},
  {Neilsen}, {Nichol}, {Nord}, {Ogando}, {Peoples}, {Petravick}, {Plazas},
  {Romer}, {Roodman}, {Sako}, {Sanchez}, {Scarpine}, {Schubnell}, {Smith},
  {Soares-Santos}, {Sobreira}, {Suchyta}, {Swanson}, {Tarle}, {Thaler},
  {Thomas}, {Wester}, {Zuntz}, and {DES Collaboration}}]{bechtol15}
{Bechtol} K, {Drlica-Wagner} A, {Balbinot} E, et~al (2015) {Eight New Milky Way
  Companions Discovered in First-year Dark Energy Survey Data}. \apj 807:50.
  \doi{10.1088/0004-637X/807/1/50},
  {\href{https://arxiv.org/abs/1503.02584}{{arXiv:1503.02584}}} {[astro-ph.GA]}

\bibitem[{{Behroozi} et~al(2010){Behroozi}, {Conroy}, and
  {Wechsler}}]{behroozi10}
{Behroozi} PS, {Conroy} C, {Wechsler} RH (2010) {A Comprehensive Analysis of
  Uncertainties Affecting the Stellar Mass-Halo Mass Relation for 0 < z < 4}.
  \apj 717(1):379--403. \doi{10.1088/0004-637X/717/1/379},
  {\href{https://arxiv.org/abs/1001.0015}{{arXiv:1001.0015}}} {[astro-ph.CO]}

\bibitem[{{Belokurov} and {Koposov}(2016)}]{belokurov16}
{Belokurov} V, {Koposov} SE (2016) {Stellar streams around the Magellanic
  Clouds}. \mnras 456(1):602--616. \doi{10.1093/mnras/stv2688},
  {\href{https://arxiv.org/abs/1511.03667}{{arXiv:1511.03667}}} {[astro-ph.GA]}

\bibitem[{{Belokurov} et~al(2006){Belokurov}, {Zucker}, {Evans}, {Gilmore},
  {Vidrih}, {Bramich}, {Newberg}, {Wyse}, {Irwin}, {Fellhauer}, {Hewett},
  {Walton}, {Wilkinson}, {Cole}, {Yanny}, {Rockosi}, {Beers}, {Bell},
  {Brinkmann}, {Ivezi{\'c}}, and {Lupton}}]{belokurov06}
{Belokurov} V, {Zucker} DB, {Evans} NW, et~al (2006) {The Field of Streams:
  Sagittarius and Its Siblings}. \apjl 642(2):L137--L140. \doi{10.1086/504797},
  {\href{https://arxiv.org/abs/astro-ph/0605025}{{arXiv:astro-ph/0605025}}}
  {[astro-ph]}

\bibitem[{{Besla} et~al(2007){Besla}, {Kallivayalil}, {Hernquist}, {Robertson},
  {Cox}, {van der Marel}, and {Alcock}}]{besla07}
{Besla} G, {Kallivayalil} N, {Hernquist} L, et~al (2007) {Are the Magellanic
  Clouds on Their First Passage about the Milky Way?} \apj 668:949--967.
  \doi{10.1086/521385},
  {\href{https://arxiv.org/abs/astro-ph/0703196}{{arXiv:astro-ph/0703196}}}
  {[astro-ph]}

\bibitem[{{Besla} et~al(2010){Besla}, {Kallivayalil}, {Hernquist}, {van der
  Marel}, {Cox}, and {Kere{\v{s}}}}]{besla10}
{Besla} G, {Kallivayalil} N, {Hernquist} L, et~al (2010) {Simulations of the
  Magellanic Stream in a First Infall Scenario}. \apj 721:L97--L101.
  \doi{10.1088/2041-8205/721/2/L97},
  {\href{https://arxiv.org/abs/1008.2210}{{arXiv:1008.2210}}} {[astro-ph.GA]}

\bibitem[{{Besla} et~al(2012){Besla}, {Kallivayalil}, {Hernquist}, {van der
  Marel}, {Cox}, and {Kere{\v{s}}}}]{besla12}
{Besla} G, {Kallivayalil} N, {Hernquist} L, et~al (2012) {The role of dwarf
  galaxy interactions in shaping the Magellanic System and implications for
  Magellanic Irregulars}. \mnras 421:2109--2138.
  \doi{10.1111/j.1365-2966.2012.20466.x},
  {\href{https://arxiv.org/abs/1201.1299}{{arXiv:1201.1299}}} {[astro-ph.GA]}

\bibitem[{{Bland-Hawthorn} and {Cohen}(2003)}]{bland-hawthorn03}
{Bland-Hawthorn} J, {Cohen} M (2003) {The Large-Scale Bipolar Wind in the
  Galactic Center}. \apj 582(1):246--256. \doi{10.1086/344573},
  {\href{https://arxiv.org/abs/astro-ph/0208553}{{arXiv:astro-ph/0208553}}}
  {[astro-ph]}

\bibitem[{{Bland-Hawthorn} and {Gerhard}(2016)}]{bland-hawthorn16}
{Bland-Hawthorn} J, {Gerhard} O (2016) {The Galaxy in Context: Structural,
  Kinematic, and Integrated Properties}. \araa 54:529--596.
  \doi{10.1146/annurev-astro-081915-023441},
  {\href{https://arxiv.org/abs/1602.07702}{{arXiv:1602.07702}}} {[astro-ph.GA]}

\bibitem[{{Bland-Hawthorn} et~al(2007){Bland-Hawthorn}, {Sutherland}, {Agertz},
  and {Moore}}]{bland-hawthorn07}
{Bland-Hawthorn} J, {Sutherland} R, {Agertz} O, et~al (2007) {The Source of
  Ionization along the Magellanic Stream}. \apj 670:L109--L112.
  \doi{10.1086/524657},
  {\href{https://arxiv.org/abs/0711.0247}{{arXiv:0711.0247}}} {[astro-ph]}

\bibitem[{{Bland-Hawthorn} et~al(2013){Bland-Hawthorn}, {Maloney},
  {Sutherland}, and {Madsen}}]{bland-hawthorn13}
{Bland-Hawthorn} J, {Maloney} PR, {Sutherland} RS, et~al (2013) {Fossil Imprint
  of a Powerful Flare at the Galactic Center along the Magellanic Stream}. \apj
  778(1):58. \doi{10.1088/0004-637X/778/1/58},
  {\href{https://arxiv.org/abs/1309.5455}{{arXiv:1309.5455}}} {[astro-ph.GA]}

\bibitem[{{Bland-Hawthorn} et~al(2019){Bland-Hawthorn}, {Maloney},
  {Sutherland}, {Groves}, {Guglielmo}, {Li}, {Curzons}, {Cecil}, and
  {Fox}}]{bland-hawthorn19}
{Bland-Hawthorn} J, {Maloney} PR, {Sutherland} R, et~al (2019) {The Large-scale
  Ionization Cones in the Galaxy}. \apj 886(1):45.
  \doi{10.3847/1538-4357/ab44c8},
  {\href{https://arxiv.org/abs/1910.02225}{{arXiv:1910.02225}}} {[astro-ph.GA]}

\bibitem[{{Bok}(1966)}]{bok66}
{Bok} BJ (1966) {Magellanic Clouds}. \araa 4:95.
  \doi{10.1146/annurev.aa.04.090166.000523}

\bibitem[{{Bregman}(1979)}]{bregman79}
{Bregman} JN (1979) {Galactic wakes and the Magellanic Stream.} \apj
  229:514--523. \doi{10.1086/156984}

\bibitem[{{Br{\"u}ns} et~al(2005){Br{\"u}ns}, {Kerp}, {Staveley-Smith},
  {Mebold}, {Putman}, {Haynes}, {Kalberla}, {Muller}, and
  {Filipovic}}]{bruns05}
{Br{\"u}ns} C, {Kerp} J, {Staveley-Smith} L, et~al (2005) {The Parkes H I
  Survey of the Magellanic System}. \aap 432:45--67.
  \doi{10.1051/0004-6361:20040321},
  {\href{https://arxiv.org/abs/astro-ph/0411453}{{arXiv:astro-ph/0411453}}}
  {[astro-ph]}

\bibitem[{{Bustard} and {Gronke}(2022)}]{bustard22}
{Bustard} C, {Gronke} M (2022) {Radiative Turbulent Mixing Layers and the
  Survival of Magellanic Debris}. \apj 933(2):120.
  \doi{10.3847/1538-4357/ac752b},
  {\href{https://arxiv.org/abs/2108.08310}{{arXiv:2108.08310}}} {[astro-ph.GA]}

\bibitem[{{Carr} et~al(2024){Carr}, {Bryan}, {Garavito-Camargo}, {Besla},
  {Setton}, and {Johnston}}]{carr24}
{Carr} C, {Bryan} GL, {Garavito-Camargo} N, et~al (2024) {The All-Sky Impact of
  the LMC on the Milky Way Circumgalactic Medium}. arXiv e-prints
  arXiv:2408.10358. \doi{10.48550/arXiv.2408.10358},
  {\href{https://arxiv.org/abs/2408.10358}{{arXiv:2408.10358}}} {[astro-ph.GA]}

\bibitem[{{Casetti-Dinescu} et~al(2014){Casetti-Dinescu}, {Moni Bidin},
  {Girard}, {M{\'e}ndez}, {Vieira}, {Korchagin}, and {van
  Altena}}]{casetti-dinescu14}
{Casetti-Dinescu} DI, {Moni Bidin} C, {Girard} TM, et~al (2014) {Recent Star
  Formation in the Leading Arm of the Magellanic Stream}. \apj 784:L37.
  \doi{10.1088/2041-8205/784/2/L37},
  {\href{https://arxiv.org/abs/1403.0517}{{arXiv:1403.0517}}} {[astro-ph.GA]}

\bibitem[{{Chamberlain} et~al(2023){Chamberlain}, {Price-Whelan}, {Besla},
  {Cunningham}, {Garavito-Camargo}, {Pe{\~n}arrubia}, and
  {Petersen}}]{chamberlain23}
{Chamberlain} K, {Price-Whelan} AM, {Besla} G, et~al (2023) {Implications of
  the Milky Way Travel Velocity for Dynamical Mass Estimates of the Local
  Group}. \apj 942(1):18. \doi{10.3847/1538-4357/aca01f},
  {\href{https://arxiv.org/abs/2204.07173}{{arXiv:2204.07173}}} {[astro-ph.GA]}

\bibitem[{{Chandra} et~al(2023){Chandra}, {Naidu}, {Conroy}, {Bonaca},
  {Zaritsky}, {Cargile}, {Caldwell}, {Johnson}, {Han}, and {Ting}}]{chandra23}
{Chandra} V, {Naidu} RP, {Conroy} C, et~al (2023) {Discovery of the Magellanic
  Stellar Stream Out to 100 kpc}. \apj 956(2):110.
  \doi{10.3847/1538-4357/acf7bf},
  {\href{https://arxiv.org/abs/2306.15719}{{arXiv:2306.15719}}} {[astro-ph.GA]}

\bibitem[{{Chandrasekhar}(1943)}]{chandrasekhar43}
{Chandrasekhar} S (1943) {Dynamical Friction. I. General Considerations: the
  Coefficient of Dynamical Friction.} \apj 97:255. \doi{10.1086/144517}

\bibitem[{{Cheng} et~al(2022){Cheng}, {Choi}, {Olsen}, {Nidever}, {Majewski},
  {Monachesi}, {Besla}, {Mu{\~n}oz Gonzalez}, {Anguiano}, {Almeida},
  {Mu{\~n}oz}, {Lane}, and {Nitschelm}}]{cheng22}
{Cheng} X, {Choi} Y, {Olsen} K, et~al (2022) {Kinematical Analysis of
  Substructure in the Southern Periphery of the Large Magellanic Cloud}. \apj
  928(1):95. \doi{10.3847/1538-4357/ac5621},
  {\href{https://arxiv.org/abs/2202.12789}{{arXiv:2202.12789}}} {[astro-ph.GA]}

\bibitem[{{Choi} et~al(2022){Choi}, {Olsen}, {Besla}, {van der Marel},
  {Zivick}, {Kallivayalil}, and {Nidever}}]{choi22}
{Choi} Y, {Olsen} KAG, {Besla} G, et~al (2022) {The recent LMC-SMC collision:
  Timing and impact parameter constraints from comparison of Gaia LMC disk
  kinematics and N-body simulations}. arXiv e-prints arXiv:2201.04648.
  {\href{https://arxiv.org/abs/2201.04648}{{arXiv:2201.04648}}} {[astro-ph.GA]}

\bibitem[{{Cioni} et~al(2000){Cioni}, {Habing}, and {Israel}}]{cioni00}
{Cioni} MRL, {Habing} HJ, {Israel} FP (2000) {The morphology of the Magellanic
  Clouds revealed by stars of different age: results from the DENIS survey}.
  \aap 358:L9--L12. \doi{10.48550/arXiv.astro-ph/0005057},
  {\href{https://arxiv.org/abs/astro-ph/0005057}{{arXiv:astro-ph/0005057}}}
  {[astro-ph]}

\bibitem[{{Conroy} et~al(2021){Conroy}, {Naidu}, {Garavito-Camargo}, {Besla},
  {Zaritsky}, {Bonaca}, and {Johnson}}]{conroy21}
{Conroy} C, {Naidu} RP, {Garavito-Camargo} N, et~al (2021) {All-sky dynamical
  response of the Galactic halo to the Large Magellanic Cloud}. \nat
  592(7855):534--536. \doi{10.1038/s41586-021-03385-7},
  {\href{https://arxiv.org/abs/2104.09515}{{arXiv:2104.09515}}} {[astro-ph.GA]}

\bibitem[{{Crowther} et~al(2016){Crowther}, {Caballero-Nieves}, {Bostroem},
  {Ma{\'\i}z Apell{\'a}niz}, {Schneider}, {Walborn}, {Angus}, {Brott},
  {Bonanos}, {de Koter}, {de Mink}, {Evans}, {Gr{\"a}fener}, {Herrero},
  {Howarth}, {Langer}, {Lennon}, {Puls}, {Sana}, and {Vink}}]{crowther16}
{Crowther} PA, {Caballero-Nieves} SM, {Bostroem} KA, et~al (2016) {The R136
  star cluster dissected with Hubble Space Telescope/STIS. I. Far-ultraviolet
  spectroscopic census and the origin of He II {\ensuremath{\lambda}}1640 in
  young star clusters}. \mnras 458(1):624--659. \doi{10.1093/mnras/stw273},
  {\href{https://arxiv.org/abs/1603.04994}{{arXiv:1603.04994}}} {[astro-ph.SR]}

\bibitem[{{de Grijs} et~al(2014){de Grijs}, {Wicker}, and {Bono}}]{degrijs14}
{de Grijs} R, {Wicker} JE, {Bono} G (2014) {Clustering of Local Group
  Distances: Publication Bias or Correlated Measurements? I. The Large
  Magellanic Cloud}. \aj 147(5):122. \doi{10.1088/0004-6256/147/5/122},
  {\href{https://arxiv.org/abs/1403.3141}{{arXiv:1403.3141}}} {[astro-ph.GA]}

\bibitem[{{de Vaucouleurs} and {Freeman}(1972)}]{devaucouleurs72}
{de Vaucouleurs} G, {Freeman} KC (1972) {Structure and dynamics of barred
  spiral galaxies, in particular of the Magellanic type}. Vistas in Astronomy
  14(1):163--294. \doi{10.1016/0083-6656(72)90026-8}

\bibitem[{{Dharmawardena} et~al(2024){Dharmawardena}, {Bailer-Jones},
  {Fouesneau}, {Foreman-Mackey}, {Coronica}, {Colnaghi}, {M{\"u}ller}, and
  {Wilson}}]{dharmawardena24}
{Dharmawardena} TE, {Bailer-Jones} CAL, {Fouesneau} M, et~al (2024) {All-sky
  three-dimensional dust density and extinction Maps of the Milky Way out to
  2.8 kpc}. \mnras 532(3):3480--3498. \doi{10.1093/mnras/stae1474},
  {\href{https://arxiv.org/abs/2406.06740}{{arXiv:2406.06740}}} {[astro-ph.GA]}

\bibitem[{{Di Teodoro} et~al(2019){Di Teodoro}, {McClure-Griffiths}, {Jameson},
  {D{\'e}nes}, {Dickey}, {Stanimirovi{\'c}}, {Staveley-Smith}, {Anderson},
  {Bunton}, {Chippendale}, {Lee-Waddell}, {MacLeod}, and
  {Voronkov}}]{diteodoro19}
{Di Teodoro} EM, {McClure-Griffiths} NM, {Jameson} KE, et~al (2019) {On the
  dynamics of the Small Magellanic Cloud through high-resolution ASKAP H I
  observations}. \mnras 483(1):392--406. \doi{10.1093/mnras/sty3095},
  {\href{https://arxiv.org/abs/1811.09627}{{arXiv:1811.09627}}} {[astro-ph.GA]}

\bibitem[{{Dias} et~al(2021){Dias}, {Angelo}, {Oliveira}, {Maia}, {Parisi}, {De
  Bortoli}, {Souza}, {Katime Santrich}, {Bassino}, {Barbuy}, {Bica}, {Geisler},
  {Kerber}, {P{\'e}rez-Villegas}, {Quint}, {Sanmartim}, {Santos}, and
  {Westera}}]{dias21}
{Dias} B, {Angelo} MS, {Oliveira} RAP, et~al (2021) {The VISCACHA survey. III.
  Star clusters counterpart of the Magellanic Bridge and Counter-Bridge in 8D}.
  \aap 647:L9. \doi{10.1051/0004-6361/202040015},
  {\href{https://arxiv.org/abs/2103.02600}{{arXiv:2103.02600}}} {[astro-ph.GA]}

\bibitem[{{Dias} et~al(2022){Dias}, {Parisi}, {Angelo}, {Maia}, {Oliveira},
  {Souza}, {Kerber}, {Santos}, {P{\'e}rez-Villegas}, {Sanmartim}, {Quint},
  {Fraga}, {Barbuy}, {Bica}, {Santrich}, {Hernandez-Jimenez}, {Geisler},
  {Minniti}, {De B{\'o}rtoli}, {Bassino}, and {Rocha}}]{dias22}
{Dias} B, {Parisi} MC, {Angelo} M, et~al (2022) {The VISCACHA survey - IV. The
  SMC West Halo in 8D}. \mnras 512(3):4334--4351. \doi{10.1093/mnras/stac259},
  {\href{https://arxiv.org/abs/2201.11119}{{arXiv:2201.11119}}} {[astro-ph.GA]}

\bibitem[{{Diaz} and {Bekki}(2011)}]{diaz11}
{Diaz} J, {Bekki} K (2011) {Constraining the orbital history of the Magellanic
  Clouds: a new bound scenario suggested by the tidal origin of the Magellanic
  Stream}. \mnras 413(3):2015--2020. \doi{10.1111/j.1365-2966.2011.18289.x},
  {\href{https://arxiv.org/abs/1101.2500}{{arXiv:1101.2500}}} {[astro-ph.GA]}

\bibitem[{{Diaz} and {Bekki}(2012)}]{diaz12}
{Diaz} JD, {Bekki} K (2012) {The Tidal Origin of the Magellanic Stream and the
  Possibility of a Stellar Counterpart}. \apj 750:36.
  \doi{10.1088/0004-637X/750/1/36},
  {\href{https://arxiv.org/abs/1112.6191}{{arXiv:1112.6191}}} {[astro-ph.GA]}

\bibitem[{{D'Onghia} and {Fox}(2016)}]{donghia16}
{D'Onghia} E, {Fox} AJ (2016) {The Magellanic Stream: Circumnavigating the
  Galaxy}. \annrev 54:363--400. \doi{10.1146/annurev-astro-081915-023251},
  {\href{https://arxiv.org/abs/1511.05853}{{arXiv:1511.05853}}} {[astro-ph.GA]}

\bibitem[{{D'Onghia} and {Lake}(2008)}]{donghia08}
{D'Onghia} E, {Lake} G (2008) {Small Dwarf Galaxies within Larger Dwarfs: Why
  Some Are Luminous while Most Go Dark}. \apj 686:L61. \doi{10.1086/592995},
  {\href{https://arxiv.org/abs/0802.0001}{{arXiv:0802.0001}}} {[astro-ph]}

\bibitem[{{Edenhofer} et~al(2024){Edenhofer}, {Zucker}, {Frank}, {Saydjari},
  {Speagle}, {Finkbeiner}, and {En{\ss}lin}}]{edenhofer24}
{Edenhofer} G, {Zucker} C, {Frank} P, et~al (2024) {A parsec-scale Galactic 3D
  dust map out to 1.25 kpc from the Sun}. \aap 685:A82.
  \doi{10.1051/0004-6361/202347628},
  {\href{https://arxiv.org/abs/2308.01295}{{arXiv:2308.01295}}} {[astro-ph.GA]}

\bibitem[{{El Youssoufi} et~al(2023){El Youssoufi}, {Cioni}, {Kacharov},
  {Bell}, {Matjevi{\'c}}, {Bekki}, {de Grijs}, {Ivanov}, and {van
  Loon}}]{elyoussoufi23}
{El Youssoufi} D, {Cioni} MRL, {Kacharov} N, et~al (2023) {Kinematics of
  stellar substructures in the small magellanic cloud}. \mnras 523(1):347--364.
  \doi{10.1093/mnras/stad1339},
  {\href{https://arxiv.org/abs/2304.14368}{{arXiv:2304.14368}}} {[astro-ph.GA]}

\bibitem[{{Erkal} and {Belokurov}(2020)}]{erkal20}
{Erkal} D, {Belokurov} VA (2020) {Limit on the LMC mass from a census of its
  satellites}. \mnras 495(3):2554--2563. \doi{10.1093/mnras/staa1238},
  {\href{https://arxiv.org/abs/1907.09484}{{arXiv:1907.09484}}} {[astro-ph.GA]}

\bibitem[{{Erkal} et~al(2019){Erkal}, {Belokurov}, {Laporte}, {Koposov}, {Li},
  {Grillmair}, {Kallivayalil}, {Price-Whelan}, {Evans}, {Hawkins}, {Hendel},
  {Mateu}, {Navarro}, {del Pino}, {Slater}, {Sohn}, and {Orphan Aspen Treasury
  Collaboration}}]{erkal19}
{Erkal} D, {Belokurov} V, {Laporte} CFP, et~al (2019) {The total mass of the
  Large Magellanic Cloud from its perturbation on the Orphan stream}. \mnras
  487(2):2685--2700. \doi{10.1093/mnras/stz1371},
  {\href{https://arxiv.org/abs/1812.08192}{{arXiv:1812.08192}}} {[astro-ph.GA]}

\bibitem[{{For} et~al(2013){For}, {Staveley-Smith}, and
  {McClure-Griffiths}}]{for13}
{For} BQ, {Staveley-Smith} L, {McClure-Griffiths} NM (2013) {Galactic All-Sky
  Survey High-velocity Clouds in the Region of the Magellanic Leading Arm}.
  \apj 764(1):74. \doi{10.1088/0004-637X/764/1/74},
  {\href{https://arxiv.org/abs/1208.5583}{{arXiv:1208.5583}}} {[astro-ph.GA]}

\bibitem[{{Fox} et~al(2013){Fox}, {Richter}, {Wakker}, {Lehner}, {Howk}, {Ben
  Bekhti}, {Bland-Hawthorn}, and {Lucas}}]{fox13}
{Fox} AJ, {Richter} P, {Wakker} BP, et~al (2013) {The COS/UVES Absorption
  Survey of the Magellanic Stream. I. One-tenth Solar Abundances along the Body
  of the Stream}. \apj 772:110. \doi{10.1088/0004-637X/772/2/110},
  {\href{https://arxiv.org/abs/1304.4240}{{arXiv:1304.4240}}} {[astro-ph.GA]}

\bibitem[{{Fox} et~al(2014){Fox}, {Wakker}, {Barger}, {Hernandez}, {Richter},
  {Lehner}, {Bland-Hawthorn}, {Charlton}, {Westmeier}, {Thom}, {Tumlinson},
  {Misawa}, {Howk}, {Haffner}, {Ely}, {Rodriguez-Hidalgo}, and
  {Kumari}}]{fox14}
{Fox} AJ, {Wakker} BP, {Barger} KA, et~al (2014) {The COS/UVES Absorption
  Survey of the Magellanic Stream. III. Ionization, Total Mass, and Inflow Rate
  onto the Milky Way}. \apj 787:147. \doi{10.1088/0004-637X/787/2/147},
  {\href{https://arxiv.org/abs/1404.5514}{{arXiv:1404.5514}}} {[astro-ph.GA]}

\bibitem[{{Fox} et~al(2018){Fox}, {Barger}, {Wakker}, {Richter}, {Antwi-Danso},
  {Casetti-Dinescu}, {Howk}, {Lehner}, {D'Onghia}, {Crowther}, and
  {Lockman}}]{fox18}
{Fox} AJ, {Barger} KA, {Wakker} BP, et~al (2018) {Chemical Abundances in the
  Leading Arm of the Magellanic Stream}. \apj 854:142.
  \doi{10.3847/1538-4357/aaa9bb},
  {\href{https://arxiv.org/abs/1801.06446}{{arXiv:1801.06446}}} {[astro-ph.GA]}

\bibitem[{{Fujimoto} and {Sofue}(1976)}]{fujimoto76}
{Fujimoto} M, {Sofue} Y (1976) {Dynamical evolution of the triple system of the
  Galaxy, the Large and Small Magellanic Clouds.} \aap 47:263--291

\bibitem[{{Gaia Collaboration} et~al(2016){Gaia Collaboration}, {Prusti}, {de
  Bruijne}, {Brown}, {Vallenari}, {Babusiaux}, {Bailer-Jones}, {Bastian},
  {Biermann}, {Evans}, {Eyer}, {Jansen}, {Jordi}, {Klioner}, {Lammers},
  {Lindegren}, {Luri}, {Mignard}, {Milligan}, {Panem}, {Poinsignon},
  {Pourbaix}, {Randich}, {Sarri}, {Sartoretti}, {Siddiqui}, {Soubiran},
  {Valette}, {van Leeuwen}, {Walton}, {Aerts}, {Arenou}, {Cropper}, {Drimmel},
  {H{\o}g}, {Katz}, {Lattanzi}, {O'Mullane}, {Grebel}, {Holland}, {Huc},
  {Passot}, {Bramante}, {Cacciari}, {Casta{\~n}eda}, {Chaoul}, {Cheek}, {De
  Angeli}, {Fabricius}, {Guerra}, {Hern{\'a}ndez}, {Jean-Antoine-Piccolo},
  {Masana}, {Messineo}, {Mowlavi}, {Nienartowicz}, {Ord{\'o}{\~n}ez-Blanco},
  {Panuzzo}, {Portell}, {Richards}, {Riello}, {Seabroke}, {Tanga},
  {Th{\'e}venin}, {Torra}, {Els}, {Gracia-Abril}, {Comoretto},
  {Garcia-Reinaldos}, {Lock}, {Mercier}, {Altmann}, {Andrae}, {Astraatmadja},
  {Bellas-Velidis}, {Benson}, {Berthier}, {Blomme}, {Busso}, {Carry},
  {Cellino}, {Clementini}, {Cowell}, {Creevey}, {Cuypers}, {Davidson}, {De
  Ridder}, {de Torres}, {Delchambre}, {Dell'Oro}, {Ducourant}, {Fr{\'e}mat},
  {Garc{\'\i}a-Torres}, {Gosset}, {Halbwachs}, {Hambly}, {Harrison}, {Hauser},
  {Hestroffer}, {Hodgkin}, {Huckle}, {Hutton}, {Jasniewicz}, {Jordan},
  {Kontizas}, {Korn}, {Lanzafame}, {Manteiga}, {Moitinho}, {Muinonen},
  {Osinde}, {Pancino}, {Pauwels}, {Petit}, {Recio-Blanco}, {Robin}, {Sarro},
  {Siopis}, {Smith}, {Smith}, {Sozzetti}, {Thuillot}, {van Reeven}, {Viala},
  {Abbas}, {Abreu Aramburu}, {Accart}, {Aguado}, {Allan}, {Allasia},
  {Altavilla}, {{\'A}lvarez}, {Alves}, {Anderson}, {Andrei}, {Anglada Varela},
  {Antiche}, {Antoja}, {Ant{\'o}n}, {Arcay}, {Atzei}, {Ayache}, {Bach},
  {Baker}, {Balaguer-N{\'u}{\~n}ez}, {Barache}, {Barata}, {Barbier}, {Barblan},
  {Baroni}, {Barrado y Navascu{\'e}s}, {Barros}, {Barstow}, {Becciani},
  {Bellazzini}, {Bellei}, {Bello Garc{\'\i}a}, {Belokurov}, {Bendjoya},
  {Berihuete}, {Bianchi}, {Bienaym{\'e}}, {Billebaud}, {Blagorodnova},
  {Blanco-Cuaresma}, {Boch}, {Bombrun}, {Borrachero}, {Bouquillon}, {Bourda},
  {Bouy}, {Bragaglia}, {Breddels}, {Brouillet}, {Br{\"u}semeister},
  {Bucciarelli}, {Budnik}, {Burgess}, {Burgon}, {Burlacu}, {Busonero}, {Buzzi},
  {Caffau}, {Cambras}, {Campbell}, {Cancelliere}, {Cantat-Gaudin}, {Carlucci},
  {Carrasco}, {Castellani}, {Charlot}, {Charnas}, {Charvet}, {Chassat},
  {Chiavassa}, {Clotet}, {Cocozza}, {Collins}, {Collins}, {Costigan}, {Crifo},
  {Cross}, {Crosta}, {Crowley}, {Dafonte}, {Damerdji}, {Dapergolas}, {David},
  {David}, {De Cat}, {de Felice}, {de Laverny}, {De Luise}, {De March}, {de
  Martino}, {de Souza}, {Debosscher}, {del Pozo}, {Delbo}, {Delgado},
  {Delgado}, {di Marco}, {Di Matteo}, {Diakite}, {Distefano}, {Dolding}, {Dos
  Anjos}, {Drazinos}, {Dur{\'a}n}, {Dzigan}, {Ecale}, {Edvardsson}, {Enke},
  {Erdmann}, {Escolar}, {Espina}, {Evans}, {Eynard Bontemps}, {Fabre},
  {Fabrizio}, {Faigler}, {Falc{\~a}o}, {Farr{\`a}s Casas}, {Faye}, {Federici},
  {Fedorets}, {Fern{\'a}ndez-Hern{\'a}ndez}, {Fernique}, {Fienga}, {Figueras},
  {Filippi}, {Findeisen}, {Fonti}, {Fouesneau}, {Fraile}, {Fraser}, {Fuchs},
  {Furnell}, {Gai}, {Galleti}, {Galluccio}, {Garabato}, {Garc{\'\i}a-Sedano},
  {Gar{\'e}}, {Garofalo}, {Garralda}, {Gavras}, {Gerssen}, {Geyer}, {Gilmore},
  {Girona}, {Giuffrida}, {Gomes}, {Gonz{\'a}lez-Marcos},
  {Gonz{\'a}lez-N{\'u}{\~n}ez}, {Gonz{\'a}lez-Vidal}, {Granvik}, {Guerrier},
  {Guillout}, {Guiraud}, {G{\'u}rpide}, {Guti{\'e}rrez-S{\'a}nchez}, {Guy},
  {Haigron}, {Hatzidimitriou}, {Haywood}, {Heiter}, {Helmi}, {Hobbs},
  {Hofmann}, {Holl}, {Holland}, {Hunt}, {Hypki}, {Icardi}, {Irwin}, {Jevardat
  de Fombelle}, {Jofr{\'e}}, {Jonker}, {Jorissen}, {Julbe}, {Karampelas},
  {Kochoska}, {Kohley}, {Kolenberg}, {Kontizas}, {Koposov}, {Kordopatis},
  {Koubsky}, {Kowalczyk}, {Krone-Martins}, {Kudryashova}, {Kull}, {Bachchan},
  {Lacoste-Seris}, {Lanza}, {Lavigne}, {Le Poncin-Lafitte}, {Lebreton},
  {Lebzelter}, {Leccia}, {Leclerc}, {Lecoeur-Taibi}, {Lemaitre}, {Lenhardt},
  {Leroux}, {Liao}, {Licata}, {Lindstr{\o}m}, {Lister}, {Livanou}, {Lobel},
  {L{\"o}ffler}, {L{\'o}pez}, {Lopez-Lozano}, {Lorenz}, {Loureiro},
  {MacDonald}, {Magalh{\~a}es Fernandes}, {Managau}, {Mann}, {Mantelet},
  {Marchal}, {Marchant}, {Marconi}, {Marie}, {Marinoni}, {Marrese},
  {Marschalk{\'o}}, {Marshall}, {Mart{\'\i}n-Fleitas}, {Martino}, {Mary},
  {Matijevi{\v{c}}}, {Mazeh}, {McMillan}, {Messina}, {Mestre}, {Michalik},
  {Millar}, {Miranda}, {Molina}, {Molinaro}, {Molinaro}, {Moln{\'a}r},
  {Moniez}, {Montegriffo}, {Monteiro}, {Mor}, {Mora}, {Morbidelli}, {Morel},
  {Morgenthaler}, {Morley}, {Morris}, {Mulone}, {Muraveva}, {Musella},
  {Narbonne}, {Nelemans}, {Nicastro}, {Noval}, {Ord{\'e}novic},
  {Ordieres-Mer{\'e}}, {Osborne}, {Pagani}, {Pagano}, {Pailler}, {Palacin},
  {Palaversa}, {Parsons}, {Paulsen}, {Pecoraro}, {Pedrosa}, {Pentik{\"a}inen},
  {Pereira}, {Pichon}, {Piersimoni}, {Pineau}, {Plachy}, {Plum}, {Poujoulet},
  {Pr{\v{s}}a}, {Pulone}, {Ragaini}, {Rago}, {Rambaux}, {Ramos-Lerate},
  {Ranalli}, {Rauw}, {Read}, {Regibo}, {Renk}, {Reyl{\'e}}, {Ribeiro},
  {Rimoldini}, {Ripepi}, {Riva}, {Rixon}, {Roelens}, {Romero-G{\'o}mez},
  {Rowell}, {Royer}, {Rudolph}, {Ruiz-Dern}, {Sadowski}, {Sagrist{\`a}
  Sell{\'e}s}, {Sahlmann}, {Salgado}, {Salguero}, {Sarasso}, {Savietto},
  {Schnorhk}, {Schultheis}, {Sciacca}, {Segol}, {Segovia}, {Segransan},
  {Serpell}, {Shih}, {Smareglia}, {Smart}, {Smith}, {Solano}, {Solitro},
  {Sordo}, {Soria Nieto}, {Souchay}, {Spagna}, {Spoto}, {Stampa}, {Steele},
  {Steidelm{\"u}ller}, {Stephenson}, {Stoev}, {Suess}, {S{\"u}veges}, {Surdej},
  {Szabados}, {Szegedi-Elek}, {Tapiador}, {Taris}, {Tauran}, {Taylor},
  {Teixeira}, {Terrett}, {Tingley}, {Trager}, {Turon}, {Ulla}, {Utrilla},
  {Valentini}, {van Elteren}, {Van Hemelryck}, {van Leeuwen}, {Varadi},
  {Vecchiato}, {Veljanoski}, {Via}, {Vicente}, {Vogt}, {Voss}, {Votruba},
  {Voutsinas}, {Walmsley}, {Weiler}, {Weingrill}, {Werner}, {Wevers},
  {Whitehead}, {Wyrzykowski}, {Yoldas}, {{\v{Z}}erjal}, {Zucker}, {Zurbach},
  {Zwitter}, {Alecu}, {Allen}, {Allende Prieto}, {Amorim},
  {Anglada-Escud{\'e}}, {Arsenijevic}, {Azaz}, {Balm}, {Beck}, {Bernstein},
  {Bigot}, {Bijaoui}, {Blasco}, {Bonfigli}, {Bono}, {Boudreault}, {Bressan},
  {Brown}, {Brunet}, {Bunclark}, {Buonanno}, {Butkevich}, {Carret}, {Carrion},
  {Chemin}, {Ch{\'e}reau}, {Corcione}, {Darmigny}, {de Boer}, {de Teodoro}, {de
  Zeeuw}, {Delle Luche}, {Domingues}, {Dubath}, {Fodor}, {Fr{\'e}zouls},
  {Fries}, {Fustes}, {Fyfe}, {Gallardo}, {Gallegos}, {Gardiol}, {Gebran},
  {Gomboc}, {G{\'o}mez}, {Grux}, {Gueguen}, {Heyrovsky}, {Hoar}, {Iannicola},
  {Isasi Parache}, {Janotto}, {Joliet}, {Jonckheere}, {Keil}, {Kim},
  {Klagyivik}, {Klar}, {Knude}, {Kochukhov}, {Kolka}, {Kos}, {Kutka}, {Lainey},
  {LeBouquin}, {Liu}, {Loreggia}, {Makarov}, {Marseille}, {Martayan},
  {Martinez-Rubi}, {Massart}, {Meynadier}, {Mignot}, {Munari}, {Nguyen},
  {Nordlander}, {Ocvirk}, {O'Flaherty}, {Olias Sanz}, {Ortiz}, {Osorio},
  {Oszkiewicz}, {Ouzounis}, {Palmer}, {Park}, {Pasquato}, {Peltzer}, {Peralta},
  {P{\'e}turaud}, {Pieniluoma}, {Pigozzi}, {Poels}, {Prat}, {Prod'homme},
  {Raison}, {Rebordao}, {Risquez}, {Rocca-Volmerange}, {Rosen}, {Ruiz-Fuertes},
  {Russo}, {Sembay}, {Serraller Vizcaino}, {Short}, {Siebert}, {Silva},
  {Sinachopoulos}, {Slezak}, {Soffel}, {Sosnowska}, {Strai{\v{z}}ys}, {ter
  Linden}, {Terrell}, {Theil}, {Tiede}, {Troisi}, {Tsalmantza}, {Tur},
  {Vaccari}, {Vachier}, {Valles}, {Van Hamme}, {Veltz}, {Virtanen}, {Wallut},
  {Wichmann}, {Wilkinson}, {Ziaeepour}, and {Zschocke}}]{gaia16}
{Gaia Collaboration}, {Prusti} T, {de Bruijne} JHJ, et~al (2016) {The Gaia
  mission}. \aap 595:A1. \doi{10.1051/0004-6361/201629272},
  {\href{https://arxiv.org/abs/1609.04153}{{arXiv:1609.04153}}} {[astro-ph.IM]}

\bibitem[{{Garavito-Camargo} et~al(2019){Garavito-Camargo}, {Besla}, {Laporte},
  {Johnston}, {G{\'o}mez}, and {Watkins}}]{garavito-camargo19}
{Garavito-Camargo} N, {Besla} G, {Laporte} CFP, et~al (2019) {Hunting for the
  Dark Matter Wake Induced by the Large Magellanic Cloud}. \apj 884(1):51.
  \doi{10.3847/1538-4357/ab32eb},
  {\href{https://arxiv.org/abs/1902.05089}{{arXiv:1902.05089}}} {[astro-ph.GA]}

\bibitem[{{Gardiner} and {Noguchi}(1996)}]{gardiner96}
{Gardiner} LT, {Noguchi} M (1996) {N-body simulations of the Small Magellanic
  Cloud and the Magellanic Stream}. \mnras 278(1):191--208.
  \doi{10.1093/mnras/278.1.191},
  {\href{https://arxiv.org/abs/astro-ph/9503095}{{arXiv:astro-ph/9503095}}}
  {[astro-ph]}

\bibitem[{{Gibson} et~al(2000){Gibson}, {Giroux}, {Penton}, {Putman}, {Stocke},
  and {Shull}}]{gibson00}
{Gibson} BK, {Giroux} ML, {Penton} SV, et~al (2000) {Metal Abundances in the
  Magellanic Stream}. \aj 120(4):1830--1840. \doi{10.1086/301545},
  {\href{https://arxiv.org/abs/astro-ph/0007078}{{arXiv:astro-ph/0007078}}}
  {[astro-ph]}

\bibitem[{{G{\'o}mez} et~al(2015){G{\'o}mez}, {Besla}, {Carpintero},
  {Villalobos}, {O'Shea}, and {Bell}}]{gomez15}
{G{\'o}mez} FA, {Besla} G, {Carpintero} DD, et~al (2015) {And Yet it Moves: The
  Dangers of Artificially Fixing the Milky Way Center of Mass in the Presence
  of a Massive Large Magellanic Cloud}. \apj 802(2):128.
  \doi{10.1088/0004-637X/802/2/128},
  {\href{https://arxiv.org/abs/1408.4128}{{arXiv:1408.4128}}} {[astro-ph.GA]}

\bibitem[{{Graczyk} et~al(2020){Graczyk}, {Pietrzy{\'n}ski}, {Thompson},
  {Gieren}, {Zgirski}, {Villanova}, {G{\'o}rski}, {Wielg{\'o}rski},
  {Karczmarek}, {Narloch}, {Pilecki}, {Taormina}, {Smolec}, {Suchomska},
  {Gallenne}, {Nardetto}, {Storm}, {Kudritzki}, {Ka{\l}uszy{\'n}ski}, and
  {Pych}}]{graczyk20}
{Graczyk} D, {Pietrzy{\'n}ski} G, {Thompson} IB, et~al (2020) {A Distance
  Determination to the Small Magellanic Cloud with an Accuracy of Better than
  Two Percent Based on Late-type Eclipsing Binary Stars}. \apj 904(1):13.
  \doi{10.3847/1538-4357/abbb2b},
  {\href{https://arxiv.org/abs/2010.08754}{{arXiv:2010.08754}}} {[astro-ph.GA]}

\bibitem[{{Guglielmo} et~al(2014){Guglielmo}, {Lewis}, and
  {Bland-Hawthorn}}]{guglielmo14}
{Guglielmo} M, {Lewis} GF, {Bland-Hawthorn} J (2014) {A genetic approach to the
  history of the Magellanic Clouds}. \mnras 444(2):1759--1774.
  \doi{10.1093/mnras/stu1549},
  {\href{https://arxiv.org/abs/1407.8298}{{arXiv:1407.8298}}} {[astro-ph.GA]}

\bibitem[{{Guo} et~al(2010){Guo}, {White}, {Li}, and {Boylan-Kolchin}}]{guo10}
{Guo} Q, {White} S, {Li} C, et~al (2010) {How do galaxies populate dark matter
  haloes?} \mnras 404(3):1111--1120. \doi{10.1111/j.1365-2966.2010.16341.x},
  {\href{https://arxiv.org/abs/0909.4305}{{arXiv:0909.4305}}} {[astro-ph.CO]}

\bibitem[{{Hammer} et~al(2015){Hammer}, {Yang}, {Flores}, {Puech}, and
  {Fouquet}}]{hammer15}
{Hammer} F, {Yang} YB, {Flores} H, et~al (2015) {The Magellanic Stream System.
  I. Ram-Pressure Tails and the Relics of the Collision Between the Magellanic
  Clouds}. \apj 813:110. \doi{10.1088/0004-637X/813/2/110},
  {\href{https://arxiv.org/abs/1510.00096}{{arXiv:1510.00096}}} {[astro-ph.GA]}

\bibitem[{{Ibata} et~al(1994){Ibata}, {Gilmore}, and {Irwin}}]{ibata94}
{Ibata} RA, {Gilmore} G, {Irwin} MJ (1994) {A dwarf satellite galaxy in
  Sagittarius}. \nat 370(6486):194--196. \doi{10.1038/370194a0}

\bibitem[{{Irwin} et~al(1990){Irwin}, {Demers}, and {Kunkel}}]{irwin90}
{Irwin} MJ, {Demers} S, {Kunkel} WE (1990) {A Blue Stellar Link Between the
  Magellanic Clouds}. \aj 99:191. \doi{10.1086/115319}

\bibitem[{{Jones} et~al(1994){Jones}, {Klemola}, and {Lin}}]{jones94}
{Jones} BF, {Klemola} AR, {Lin} DNC (1994) {Proper Motion of The Large
  Magellanic Cloud And The Mass of The Galaxy. I. Observational Result}. \aj
  107:1333. \doi{10.1086/116947}

\bibitem[{{Kalari} et~al(2022){Kalari}, {Horch}, {Salinas}, {Vink}, {Andersen},
  {Bestenlehner}, and {Rubio}}]{kalari22}
{Kalari} VM, {Horch} EP, {Salinas} R, et~al (2022) {Resolving the Core of R136
  in the Optical}. \apj 935(2):162. \doi{10.3847/1538-4357/ac8424},
  {\href{https://arxiv.org/abs/2207.13078}{{arXiv:2207.13078}}} {[astro-ph.SR]}

\bibitem[{{Kalberla} et~al(2005){Kalberla}, {Burton}, {Hartmann}, {Arnal},
  {Bajaja}, {Morras}, and {P{\"o}ppel}}]{kalberla05}
{Kalberla} PMW, {Burton} WB, {Hartmann} D, et~al (2005) {The
  Leiden/Argentine/Bonn (LAB) Survey of Galactic HI. Final data release of the
  combined LDS and IAR surveys with improved stray-radiation corrections}. \aap
  440(2):775--782. \doi{10.1051/0004-6361:20041864},
  {\href{https://arxiv.org/abs/astro-ph/0504140}{{arXiv:astro-ph/0504140}}}
  {[astro-ph]}

\bibitem[{{Kallivayalil} et~al(2006){Kallivayalil}, {van der Marel}, {Alcock},
  {Axelrod}, {Cook}, {Drake}, and {Geha}}]{kallivayalil06}
{Kallivayalil} N, {van der Marel} RP, {Alcock} C, et~al (2006) {The Proper
  Motion of the Large Magellanic Cloud Using HST}. \apj 638:772--785.
  \doi{10.1086/498972},
  {\href{https://arxiv.org/abs/astro-ph/0508457}{{arXiv:astro-ph/0508457}}}
  {[astro-ph]}

\bibitem[{{Kallivayalil} et~al(2013){Kallivayalil}, {van der Marel}, {Besla},
  {Anderson}, and {Alcock}}]{kallivayalil13}
{Kallivayalil} N, {van der Marel} RP, {Besla} G, et~al (2013) {Third-epoch
  Magellanic Cloud Proper Motions. I. Hubble Space Telescope/WFC3 Data and
  Orbit Implications}. \apj 764(2):161. \doi{10.1088/0004-637X/764/2/161},
  {\href{https://arxiv.org/abs/1301.0832}{{arXiv:1301.0832}}} {[astro-ph.CO]}

\bibitem[{{Koposov} et~al(2023){Koposov}, {Erkal}, {Li}, {Da Costa},
  {Cullinane}, {Ji}, {Kuehn}, {Lewis}, {Pace}, {Shipp}, {Zucker},
  {Bland-Hawthorn}, {Lilleengen}, and {Martell}}]{koposov23}
{Koposov} SE, {Erkal} D, {Li} TS, et~al (2023) {S $^{5}$: Probing the Milky Way
  and Magellanic Clouds potentials with the 6-D map of the Orphan-Chenab
  stream}. \mnras \doi{10.1093/mnras/stad551},
  {\href{https://arxiv.org/abs/2211.04495}{{arXiv:2211.04495}}} {[astro-ph.GA]}

\bibitem[{{Krishnarao} et~al(2022){Krishnarao}, {Fox}, {D'Onghia}, {Wakker},
  {Cashman}, {Howk}, {Lucchini}, {French}, and {Lehner}}]{dk22}
{Krishnarao} D, {Fox} AJ, {D'Onghia} E, et~al (2022) {Observations of a
  Magellanic Corona}. \nat 609(7929):915--918.
  \doi{10.1038/s41586-022-05090-5},
  {\href{https://arxiv.org/abs/2209.15017}{{arXiv:2209.15017}}} {[astro-ph.GA]}

\bibitem[{{Kroupa} et~al(1994){Kroupa}, {R{\"o}ser}, and {Bastian}}]{kroupa94}
{Kroupa} P, {R{\"o}ser} S, {Bastian} U (1994) {On the motion of the Magellanic
  Clouds}. \mnras 266:412--420. \doi{10.1093/mnras/266.2.412}

\bibitem[{{Lehner} et~al(2008){Lehner}, {Howk}, {Keenan}, and
  {Smoker}}]{lehner08}
{Lehner} N, {Howk} JC, {Keenan} FP, et~al (2008) {Metallicity and Physical
  Conditions in the Magellanic Bridge}. \apj 678(1):219--233.
  \doi{10.1086/529574},
  {\href{https://arxiv.org/abs/0801.2534}{{arXiv:0801.2534}}} {[astro-ph]}

\bibitem[{{Lehner} et~al(2022){Lehner}, {Howk}, {Marasco}, and
  {Fraternali}}]{lehner22}
{Lehner} N, {Howk} JC, {Marasco} A, et~al (2022) {Intermediate- and
  high-velocity clouds in the Milky Way - I. Covering factors and vertical
  heights}. \mnras 513(3):3228--3240. \doi{10.1093/mnras/stac987},
  {\href{https://arxiv.org/abs/2202.05848}{{arXiv:2202.05848}}} {[astro-ph.GA]}

\bibitem[{{Levinson} and {Roberts}(1981)}]{levinson81}
{Levinson} FH, {Roberts} JW.~W. (1981) {A cloud/particle model of the
  interstellar medium - Galactic spiral structure}. \apj 245:465--481.
  \doi{10.1086/158823}

\bibitem[{{Lin} and {Lynden-Bell}(1977)}]{lin77}
{Lin} DNC, {Lynden-Bell} D (1977) {Simulation of the Magellanic Stream to
  estimate the total mass of the Milky Way.} \mnras 181:59--81.
  \doi{10.1093/mnras/181.2.59}

\bibitem[{{Lin} et~al(1995){Lin}, {Jones}, and {Klemola}}]{lin95}
{Lin} DNC, {Jones} BF, {Klemola} AR (1995) {The Motion of the Magellanic
  Clouds, Origin of the Magellanic Stream, and the Mass of the Milky Way}. \apj
  439:652. \doi{10.1086/175205}

\bibitem[{{Lu} et~al(1994){Lu}, {Savage}, and {Sembach}}]{lu94}
{Lu} L, {Savage} BD, {Sembach} KR (1994) {Probing the Galactic Disk and Halo:
  Metal Abundances in the Magellanic Stream}. \apjl 437:L119.
  \doi{10.1086/187697}

\bibitem[{{Lu} et~al(1998){Lu}, {Savage}, {Sembach}, {Wakker}, {Sargent}, and
  {Oosterloo}}]{lu98}
{Lu} L, {Savage} BD, {Sembach} KR, et~al (1998) {The Metallicity and Dust
  Content of HVC 287.5+22.5+240: Evidence for a Magellanic Clouds Origin}. \aj
  115(1):162--167. \doi{10.1086/300181},
  {\href{https://arxiv.org/abs/astro-ph/9710045}{{arXiv:astro-ph/9710045}}}
  {[astro-ph]}

\bibitem[{{Lucchini} et~al(2020){Lucchini}, {D'Onghia}, {Fox}, {Bustard},
  {Bland-Hawthorn}, and {Zweibel}}]{lucchini20}
{Lucchini} S, {D'Onghia} E, {Fox} AJ, et~al (2020) {The Magellanic Corona as
  the key to the formation of the Magellanic Stream.} \nat 585:203--206.
  \doi{10.1038/s41586-020-2663-4},
  {\href{https://arxiv.org/abs/2009.04368}{{arXiv:2009.04368}}} {[astro-ph.GA]}

\bibitem[{{Lucchini} et~al(2021){Lucchini}, {D'Onghia}, and {Fox}}]{lucchini21}
{Lucchini} S, {D'Onghia} E, {Fox} AJ (2021) {The Magellanic Stream at 20 kpc: A
  New Orbital History for the Magellanic Clouds}. \apjl 921(2):L36.
  \doi{10.3847/2041-8213/ac3338},
  {\href{https://arxiv.org/abs/2110.11355}{{arXiv:2110.11355}}} {[astro-ph.GA]}

\bibitem[{{Lucchini} et~al(2024){Lucchini}, {D'Onghia}, and {Fox}}]{lucchini24}
{Lucchini} S, {D'Onghia} E, {Fox} AJ (2024) {Properties of the Magellanic
  Corona}. \apj 967(1):16. \doi{10.3847/1538-4357/ad3c3b},
  {\href{https://arxiv.org/abs/2311.16221}{{arXiv:2311.16221}}} {[astro-ph.GA]}

\bibitem[{{Lynden-Bell}(1981)}]{lynden-bell81}
{Lynden-Bell} D (1981) {The dynamical age of the local group of galaxies}. The
  Observatory 101:111--114

\bibitem[{{Mastropietro} et~al(2005){Mastropietro}, {Moore}, {Mayer},
  {Wadsley}, and {Stadel}}]{mastropietro05}
{Mastropietro} C, {Moore} B, {Mayer} L, et~al (2005) {The gravitational and
  hydrodynamical interaction between the Large Magellanic Cloud and the
  Galaxy}. \mnras 363(2):509--520. \doi{10.1111/j.1365-2966.2005.09435.x},
  {\href{https://arxiv.org/abs/astro-ph/0412312}{{arXiv:astro-ph/0412312}}}
  {[astro-ph]}

\bibitem[{{Mathewson} and {Ford}(1984)}]{mathewson84}
{Mathewson} DS, {Ford} VL (1984) {HI surveys of the Magellanic system.} In:
  {van den Bergh} S, {de Boer} KSD (eds) Structure and Evolution of the
  Magellanic Clouds, pp 125--136, \doi{10.1017/S0074180900040092}

\bibitem[{{Mathewson} and {Schwarz}(1976)}]{mathewson76}
{Mathewson} DS, {Schwarz} MP (1976) {The origin of the Magellanic Stream.}
  \mnras 176:47P--51P. \doi{10.1093/mnras/176.1.47P}

\bibitem[{{Mathewson} et~al(1974){Mathewson}, {Cleary}, and
  {Murray}}]{mathewson74}
{Mathewson} DS, {Cleary} MN, {Murray} JD (1974) {The Magellanic Stream.} \apj
  190:291--296. \doi{10.1086/152875}

\bibitem[{{McClure-Griffiths} et~al(2008){McClure-Griffiths}, {Staveley-Smith},
  {Lockman}, {Calabretta}, {Ford}, {Kalberla}, {Murphy}, {Nakanishi}, and
  {Pisano}}]{mcclure-griffiths08}
{McClure-Griffiths} NM, {Staveley-Smith} L, {Lockman} FJ, et~al (2008) {An
  Interaction of a Magellanic Leading Arm High-Velocity Cloud with the Milky
  Way Disk}. \apj 673:L143. \doi{10.1086/528683},
  {\href{https://arxiv.org/abs/0712.2267}{{arXiv:0712.2267}}} {[astro-ph]}

\bibitem[{{McClure-Griffiths} et~al(2009){McClure-Griffiths}, {Pisano},
  {Calabretta}, {Ford}, {Lockman}, {Staveley-Smith}, {Kalberla}, {Bailin},
  {Dedes}, {Janowiecki}, {Gibson}, {Murphy}, {Nakanishi}, and
  {Newton-McGee}}]{mcclure-griffiths09}
{McClure-Griffiths} NM, {Pisano} DJ, {Calabretta} MR, et~al (2009) {Gass: The
  Parkes Galactic All-Sky Survey. I. Survey Description, Goals, and Initial
  Data Release}. \apjs 181(2):398--412. \doi{10.1088/0067-0049/181/2/398},
  {\href{https://arxiv.org/abs/0901.1159}{{arXiv:0901.1159}}} {[astro-ph.GA]}

\bibitem[{{Meurer} et~al(1985){Meurer}, {Bicknell}, and {Gingold}}]{meurer85}
{Meurer} GR, {Bicknell} GV, {Gingold} RA (1985) {A drag dominated model of the
  Magellanic stream.} \pasa 6(2):195--198. \doi{10.1017/S1323358000018075}

\bibitem[{{Misawa} et~al(2009){Misawa}, {Charlton}, {Kobulnicky}, {Wakker}, and
  {Bland-Hawthorn}}]{misawa09}
{Misawa} T, {Charlton} JC, {Kobulnicky} HA, et~al (2009) {The Magellanic Bridge
  as a Damped Lyman Alpha System: Physical Properties of Cold Gas Toward PKS
  0312-770}. \apj 695(2):1382--1398. \doi{10.1088/0004-637X/695/2/1382},
  {\href{https://arxiv.org/abs/0902.0208}{{arXiv:0902.0208}}} {[astro-ph.CO]}

\bibitem[{{Moore} and {Davis}(1994)}]{moore94}
{Moore} B, {Davis} M (1994) {The origin of the Magellanic Stream.} \mnras
  270:209--221. \doi{10.1093/mnras/270.2.209},
  {\href{https://arxiv.org/abs/astro-ph/9401008}{{arXiv:astro-ph/9401008}}}
  {[astro-ph]}

\bibitem[{{Murray} et~al(2019){Murray}, {Peek}, {Di Teodoro},
  {McClure-Griffiths}, {Dickey}, and {D{\'e}nes}}]{murray19}
{Murray} CE, {Peek} JEG, {Di Teodoro} EM, et~al (2019) {The 3D Kinematics of
  Gas in the Small Magellanic Cloud}. \apj 887(2):267.
  \doi{10.3847/1538-4357/ab510f},
  {\href{https://arxiv.org/abs/1910.11283}{{arXiv:1910.11283}}} {[astro-ph.GA]}

\bibitem[{{Navarrete} et~al(2023){Navarrete}, {Aguado}, {Belokurov}, {Erkal},
  {Deason}, {Cullinane}, and {Carballo-Bello}}]{navarrete23}
{Navarrete} C, {Aguado} DS, {Belokurov} V, et~al (2023) {The 3D kinematics of
  stellar substructures in the periphery of the Large Magellanic Cloud}. \mnras
  523(3):4720--4738. \doi{10.1093/mnras/stad1698},
  {\href{https://arxiv.org/abs/2302.04579}{{arXiv:2302.04579}}} {[astro-ph.GA]}

\bibitem[{{Nichols} et~al(2011){Nichols}, {Colless}, {Colless}, and
  {Bland-Hawthorn}}]{nichols11}
{Nichols} M, {Colless} J, {Colless} M, et~al (2011) {Accretion of the
  Magellanic System onto the Galaxy}. \apj 742(2):110.
  \doi{10.1088/0004-637X/742/2/110},
  {\href{https://arxiv.org/abs/1110.2784}{{arXiv:1110.2784}}} {[astro-ph.CO]}

\bibitem[{{Nidever} et~al(2008){Nidever}, {Majewski}, and {Butler
  Burton}}]{nidever08}
{Nidever} DL, {Majewski} SR, {Butler Burton} W (2008) {The Origin of the
  Magellanic Stream and Its Leading Arm}. \apj 679:432--459.
  \doi{10.1086/587042}

\bibitem[{{Nidever} et~al(2010){Nidever}, {Majewski}, {Butler Burton}, and
  {Nigra}}]{nidever10}
{Nidever} DL, {Majewski} SR, {Butler Burton} W, et~al (2010) {The
  200{\textdegree} Long Magellanic Stream System}. \apj 723:1618--1631.
  \doi{10.1088/0004-637X/723/2/1618},
  {\href{https://arxiv.org/abs/1009.0001}{{arXiv:1009.0001}}} {[astro-ph.GA]}

\bibitem[{{Odenkirchen} et~al(2003){Odenkirchen}, {Grebel}, {Dehnen}, {Rix},
  {Yanny}, {Newberg}, {Rockosi}, {Mart{\'\i}nez-Delgado}, {Brinkmann}, and
  {Pier}}]{odenkirchen03}
{Odenkirchen} M, {Grebel} EK, {Dehnen} W, et~al (2003) {The Extended Tails of
  Palomar 5: A 10{\textdegree} Arc of Globular Cluster Tidal Debris}. \aj
  126(5):2385--2407. \doi{10.1086/378601},
  {\href{https://arxiv.org/abs/astro-ph/0307446}{{arXiv:astro-ph/0307446}}}
  {[astro-ph]}

\bibitem[{{Oliveira} et~al(2023){Oliveira}, {Maia}, {Barbuy}, {Dias}, {Santos},
  {Souza}, {Kerber}, {Bica}, {Sanmartim}, {Quint}, {Fraga}, {Armond},
  {Minniti}, {Parisi}, {Katime Santrich}, {Angelo}, {P{\'e}rez-Villegas}, and
  {De B{\'o}rtoli}}]{oliveira23}
{Oliveira} RAP, {Maia} FFS, {Barbuy} B, et~al (2023) {The VISCACHA survey -
  VII. Assembly history of the Magellanic Bridge and SMC Wing from star
  clusters}. \mnras 524(2):2244--2261. \doi{10.1093/mnras/stad1827},
  {\href{https://arxiv.org/abs/2306.05503}{{arXiv:2306.05503}}} {[astro-ph.GA]}

\bibitem[{{Pardy} et~al(2018){Pardy}, {D'Onghia}, and {Fox}}]{pardy18}
{Pardy} SA, {D'Onghia} E, {Fox} AJ (2018) {Models of Tidally Induced Gas
  Filaments in the Magellanic Stream}. \apj 857:101.
  \doi{10.3847/1538-4357/aab95b},
  {\href{https://arxiv.org/abs/1802.01600}{{arXiv:1802.01600}}} {[astro-ph.GA]}

\bibitem[{{Pardy} et~al(2020){Pardy}, {D'Onghia}, {Navarro}, {Grand},
  {G{\'o}mez}, {Marinacci}, {Pakmor}, {Simpson}, and {Springel}}]{pardy20}
{Pardy} SA, {D'Onghia} E, {Navarro} JF, et~al (2020) {Satellites of Satellites:
  The Case for Carina and Fornax}. \mnras 492(2):1543--1549.
  \doi{10.1093/mnras/stz3192},
  {\href{https://arxiv.org/abs/1904.01028}{{arXiv:1904.01028}}} {[astro-ph.GA]}

\bibitem[{{Partridge} et~al(2013){Partridge}, {Lahav}, and
  {Hoffman}}]{partridge13}
{Partridge} C, {Lahav} O, {Hoffman} Y (2013) {Weighing the local group in the
  presence of dark energy.} \mnras 436:L45--L48. \doi{10.1093/mnrasl/slt109},
  {\href{https://arxiv.org/abs/1308.0970}{{arXiv:1308.0970}}} {[astro-ph.CO]}

\bibitem[{{Patel} et~al(2020){Patel}, {Kallivayalil}, {Garavito-Camargo},
  {Besla}, {Weisz}, {van der Marel}, {Boylan-Kolchin}, {Pawlowski}, and
  {G{\'o}mez}}]{patel20}
{Patel} E, {Kallivayalil} N, {Garavito-Camargo} N, et~al (2020) {The Orbital
  Histories of Magellanic Satellites Using Gaia DR2 Proper Motions}. \apj
  893(2):121. \doi{10.3847/1538-4357/ab7b75},
  {\href{https://arxiv.org/abs/2001.01746}{{arXiv:2001.01746}}} {[astro-ph.GA]}

\bibitem[{{Pe{\~n}arrubia} et~al(2016){Pe{\~n}arrubia}, {G{\'o}mez}, {Besla},
  {Erkal}, and {Ma}}]{penarrubia16}
{Pe{\~n}arrubia} J, {G{\'o}mez} FA, {Besla} G, et~al (2016) {A timing
  constraint on the (total) mass of the Large Magellanic Cloud}. \mnras
  456:L54--L58. \doi{10.1093/mnrasl/slv160},
  {\href{https://arxiv.org/abs/1507.03594}{{arXiv:1507.03594}}} {[astro-ph.GA]}

\bibitem[{{Petersen} and {Pe{\~n}arrubia}(2020)}]{petersen20}
{Petersen} MS, {Pe{\~n}arrubia} J (2020) {Reflex motion in the Milky Way
  stellar halo resulting from the Large Magellanic Cloud infall}. \mnras
  494(1):L11--L16. \doi{10.1093/mnrasl/slaa029},
  {\href{https://arxiv.org/abs/2001.09142}{{arXiv:2001.09142}}} {[astro-ph.GA]}

\bibitem[{{Petersen} and {Pe{\~n}arrubia}(2021)}]{petersen21}
{Petersen} MS, {Pe{\~n}arrubia} J (2021) {Detection of the Milky Way reflex
  motion due to the Large Magellanic Cloud infall}. Nature Astronomy
  5:251--255. \doi{10.1038/s41550-020-01254-3},
  {\href{https://arxiv.org/abs/2011.10581}{{arXiv:2011.10581}}} {[astro-ph.GA]}

\bibitem[{{Putman} et~al(1998){Putman}, {Gibson}, {Staveley-Smith}, {Banks},
  {Barnes}, {Bhatal}, {Disney}, {Ekers}, {Freeman}, {Haynes}, {Henning},
  {Jerjen}, {Kilborn}, {Koribalski}, {Knezek}, {Malin}, {Mould}, {Oosterloo},
  {Price}, {Ryder}, {Sadler}, {Stewart}, {Stootman}, {Vaile}, {Webster}, and
  {Wright}}]{putman98}
{Putman} ME, {Gibson} BK, {Staveley-Smith} L, et~al (1998) {Tidal disruption of
  the Magellanic Clouds by the Milky Way}. \nat 394(6695):752--754.
  \doi{10.1038/29466},
  {\href{https://arxiv.org/abs/astro-ph/9808023}{{arXiv:astro-ph/9808023}}}
  {[astro-ph]}

\bibitem[{{Putman} et~al(2003){Putman}, {Staveley-Smith}, {Freeman}, {Gibson},
  and {Barnes}}]{putman03}
{Putman} ME, {Staveley-Smith} L, {Freeman} KC, et~al (2003) {The Magellanic
  Stream, High-Velocity Clouds, and the Sculptor Group}. \apj 586(1):170--194.
  \doi{10.1086/344477},
  {\href{https://arxiv.org/abs/astro-ph/0209127}{{arXiv:astro-ph/0209127}}}
  {[astro-ph]}

\bibitem[{{Read} and {Erkal}(2019)}]{read19}
{Read} JI, {Erkal} D (2019) {Abundance matching with the mean star formation
  rate: there is no missing satellites problem in the Milky Way above M$_{200}$
  {\ensuremath{\sim}} {}10$^{9}$ M$_{{\ensuremath{\odot}}}$}. \mnras
  487(4):5799--5812. \doi{10.1093/mnras/stz1320},
  {\href{https://arxiv.org/abs/1807.07093}{{arXiv:1807.07093}}} {[astro-ph.GA]}

\bibitem[{{Richter} et~al(2013){Richter}, {Fox}, {Wakker}, {Lehner}, {Howk},
  {Bland -Hawthorn}, {Ben Bekhti}, and {Fechner}}]{richter13}
{Richter} P, {Fox} AJ, {Wakker} BP, et~al (2013) {The COS/UVES Absorption
  Survey of the Magellanic Stream. II. Evidence for a Complex Enrichment
  History of the Stream from the Fairall 9 Sightline}. \apj 772:111.
  \doi{10.1088/0004-637X/772/2/111},
  {\href{https://arxiv.org/abs/1304.4242}{{arXiv:1304.4242}}} {[astro-ph.GA]}

\bibitem[{{R{\r{u}}{\v{z}}i{\v{c}}ka} et~al(2010){R{\r{u}}{\v{z}}i{\v{c}}ka},
  {Theis}, and {Palou{\v{s}}}}]{ruzicka10}
{R{\r{u}}{\v{z}}i{\v{c}}ka} A, {Theis} C, {Palou{\v{s}}} J (2010) {Rotation of
  the Milky Way and the Formation of the Magellanic Stream}. \apj
  725(1):369--387. \doi{10.1088/0004-637X/725/1/369},
  {\href{https://arxiv.org/abs/1010.0942}{{arXiv:1010.0942}}} {[astro-ph.CO]}

\bibitem[{{Sandage}(1986)}]{sandage86}
{Sandage} A (1986) {The Redshift-Distance Relation. IX. Perturbation of the
  Very Nearby Velocity Field by the Mass of the Local Group}. \apj 307:1.
  \doi{10.1086/164387}

\bibitem[{{Schmidt} et~al(2020){Schmidt}, {Cioni}, {Niederhofer}, {Bekki},
  {Bell}, {de Grijs}, {Diaz}, {El Youssoufi}, {Emerson}, {Groenewegen},
  {Ivanov}, {Matijevic}, {Oliveira}, {Petr-Gotzens}, {Queiroz}, {Ripepi}, and
  {van Loon}}]{schmidt20}
{Schmidt} T, {Cioni} MRL, {Niederhofer} F, et~al (2020) {The VMC survey.
  XXXVIII. Proper motion of the Magellanic Bridge}. \aap 641:A134.
  \doi{10.1051/0004-6361/202037478},
  {\href{https://arxiv.org/abs/2006.03163}{{arXiv:2006.03163}}} {[astro-ph.GA]}

\bibitem[{{Sembach} et~al(2001){Sembach}, {Howk}, {Savage}, and
  {Shull}}]{sembach01}
{Sembach} KR, {Howk} JC, {Savage} BD, et~al (2001) {FUSE Observations of Atomic
  Abundances and Molecular Hydrogen in the Leading Arm of the Magellanic
  Stream}. \aj 121(2):992--1002. \doi{10.1086/318777},
  {\href{https://arxiv.org/abs/astro-ph/0011173}{{arXiv:astro-ph/0011173}}}
  {[astro-ph]}

\bibitem[{{Setton} et~al(2023){Setton}, {Besla}, {Patel}, {Hummels}, {Zheng},
  {Schneider}, and {Salem}}]{setton23}
{Setton} DJ, {Besla} G, {Patel} E, et~al (2023) {The Large Magellanic Cloud's
  30 kpc Bow Shock and Its Impact on the Circumgalactic Medium}. \apjl
  959(1):L11. \doi{10.3847/2041-8213/ad0da6},
  {\href{https://arxiv.org/abs/2308.10963}{{arXiv:2308.10963}}} {[astro-ph.GA]}

\bibitem[{{Shipp} et~al(2021){Shipp}, {Erkal}, {Drlica-Wagner}, {Li}, {Pace},
  {Koposov}, {Cullinane}, {Da Costa}, {Ji}, {Kuehn}, {Lewis}, {Mackey},
  {Simpson}, {Wan}, {Zucker}, {Bland-Hawthorn}, {Ferguson}, {Lilleengen}, and
  {Lilleengen}}]{shipp21}
{Shipp} N, {Erkal} D, {Drlica-Wagner} A, et~al (2021) {Measuring the Mass of
  the Large Magellanic Cloud with Stellar Streams Observed by S $^{5}$}. \apj
  923(2):149. \doi{10.3847/1538-4357/ac2e93},
  {\href{https://arxiv.org/abs/2107.13004}{{arXiv:2107.13004}}} {[astro-ph.GA]}

\bibitem[{{Stanimirovi{\'c}} et~al(2004){Stanimirovi{\'c}}, {Staveley-Smith},
  and {Jones}}]{stanimirovic04}
{Stanimirovi{\'c}} S, {Staveley-Smith} L, {Jones} PA (2004) {A New Look at the
  Kinematics of Neutral Hydrogen in the Small Magellanic Cloud}. \apj
  604(1):176--186. \doi{10.1086/381869},
  {\href{https://arxiv.org/abs/astro-ph/0312223}{{arXiv:astro-ph/0312223}}}
  {[astro-ph]}

\bibitem[{{Su} et~al(2010){Su}, {Slatyer}, and {Finkbeiner}}]{su10}
{Su} M, {Slatyer} TR, {Finkbeiner} DP (2010) {Giant Gamma-ray Bubbles from
  Fermi-LAT: Active Galactic Nucleus Activity or Bipolar Galactic Wind?} \apj
  724(2):1044--1082. \doi{10.1088/0004-637X/724/2/1044},
  {\href{https://arxiv.org/abs/1005.5480}{{arXiv:1005.5480}}} {[astro-ph.HE]}

\bibitem[{{Subramanian} and {Subramaniam}(2012)}]{subramanian12}
{Subramanian} S, {Subramaniam} A (2012) {The Three-dimensional Structure of the
  Small Magellanic Cloud}. \apj 744(2):128. \doi{10.1088/0004-637X/744/2/128},
  {\href{https://arxiv.org/abs/1109.3980}{{arXiv:1109.3980}}} {[astro-ph.CO]}

\bibitem[{{Tepper-Garc{\'i}a} et~al(2019){Tepper-Garc{\'i}a}, {Bland-Hawthorn},
  {Pawlowski}, and {Fritz}}]{tepper-garcia19}
{Tepper-Garc{\'i}a} T, {Bland-Hawthorn} J, {Pawlowski} MS, et~al (2019) {The
  Magellanic System: the puzzle of the leading gas stream}. \mnras
  488(1):918--938. \doi{10.1093/mnras/stz1659},
  {\href{https://arxiv.org/abs/1901.05636}{{arXiv:1901.05636}}} {[astro-ph.GA]}

\bibitem[{{van der Marel} and {Kallivayalil}(2014)}]{vandermarel14}
{van der Marel} RP, {Kallivayalil} N (2014) {Third-epoch Magellanic Cloud
  Proper Motions. II. The Large Magellanic Cloud Rotation Field in Three
  Dimensions}. \apj 781(2):121. \doi{10.1088/0004-637X/781/2/121},
  {\href{https://arxiv.org/abs/1305.4641}{{arXiv:1305.4641}}} {[astro-ph.CO]}

\bibitem[{{van der Marel} et~al(2002){van der Marel}, {Alves}, {Hardy}, and
  {Suntzeff}}]{vandermarel02}
{van der Marel} RP, {Alves} DR, {Hardy} E, et~al (2002) {New Understanding of
  Large Magellanic Cloud Structure, Dynamics, and Orbit from Carbon Star
  Kinematics}. \aj 124(5):2639--2663. \doi{10.1086/343775},
  {\href{https://arxiv.org/abs/astro-ph/0205161}{{arXiv:astro-ph/0205161}}}
  {[astro-ph]}

\bibitem[{{van der Marel} et~al(2009){van der Marel}, {Kallivayalil}, and
  {Besla}}]{vandermarel09}
{van der Marel} RP, {Kallivayalil} N, {Besla} G (2009) {Kinematical structure
  of the Magellanic System}. In: {Van Loon} JT, {Oliveira} JM (eds) The
  Magellanic System: Stars, Gas, and Galaxies, pp 81--92,
  \doi{10.1017/S1743921308028299}, \eprint{0809.4268}

\bibitem[{{van Kuilenburg}(1972)}]{vankuilenburg72}
{van Kuilenburg} J (1972) {A Systematic Search for High-Velocity Hydrogen
  Outside the Galactic Plane II}. \aap 16:276

\bibitem[{{Vasiliev}(2024)}]{vasiliev24}
{Vasiliev} E (2024) {Dear Magellanic Clouds, welcome back!} \mnras
  527(1):437--456. \doi{10.1093/mnras/stad2612},
  {\href{https://arxiv.org/abs/2306.04837}{{arXiv:2306.04837}}} {[astro-ph.GA]}

\bibitem[{{Vasiliev} et~al(2021){Vasiliev}, {Belokurov}, and
  {Erkal}}]{vasiliev21}
{Vasiliev} E, {Belokurov} V, {Erkal} D (2021) {Tango for three: Sagittarius,
  LMC, and the Milky Way}. \mnras 501(2):2279--2304.
  \doi{10.1093/mnras/staa3673},
  {\href{https://arxiv.org/abs/2009.10726}{{arXiv:2009.10726}}} {[astro-ph.GA]}

\bibitem[{{Vieira} et~al(2010){Vieira}, {Girard}, {van Altena}, {Zacharias},
  {Casetti-Dinescu}, {Korchagin}, {Platais}, {Monet}, {L{\'o}pez}, {Herrera},
  and {Castillo}}]{vieira10}
{Vieira} K, {Girard} TM, {van Altena} WF, et~al (2010) {Proper-motion Study of
  the Magellanic Clouds Using SPM Material}. \aj 140(6):1934--1950.
  \doi{10.1088/0004-6256/140/6/1934},
  {\href{https://arxiv.org/abs/1009.4218}{{arXiv:1009.4218}}} {[astro-ph.GA]}

\bibitem[{{Wan} et~al(2020){Wan}, {Guglielmo}, {Lewis}, {Mackey}, and
  {Ibata}}]{wan20}
{Wan} Z, {Guglielmo} M, {Lewis} GF, et~al (2020) {A SkyMapper view of the Large
  Magellanic Cloud: the dynamics of stellar populations}. \mnras
  492(1):782--795. \doi{10.1093/mnras/stz3493},
  {\href{https://arxiv.org/abs/1912.04657}{{arXiv:1912.04657}}} {[astro-ph.GA]}

\bibitem[{{Wang} et~al(2019){Wang}, {Hammer}, {Yang}, {Ripepi}, {Cioni},
  {Puech}, and {Flores}}]{wang19}
{Wang} J, {Hammer} F, {Yang} Y, et~al (2019) {Towards a complete understanding
  of the Magellanic Stream Formation}. \mnras 486(4):5907--5916.
  \doi{10.1093/mnras/stz1274},
  {\href{https://arxiv.org/abs/1905.03801}{{arXiv:1905.03801}}} {[astro-ph.GA]}

\bibitem[{{Wannier} and {Wrixon}(1972)}]{wannier72}
{Wannier} P, {Wrixon} GT (1972) {An Unusual High-Velocity Hydrogen Feature}.
  \apjl 173:L119. \doi{10.1086/180930}

\bibitem[{{Watkins} et~al(2024){Watkins}, {van der Marel}, and
  {Bennet}}]{watkins24}
{Watkins} LL, {van der Marel} RP, {Bennet} P (2024) {The Mass of the Large
  Magellanic Cloud from the Three-dimensional Kinematics of Its Globular
  Clusters}. \apj 963(2):84. \doi{10.3847/1538-4357/ad1f58},
  {\href{https://arxiv.org/abs/2401.14458}{{arXiv:2401.14458}}} {[astro-ph.GA]}

\bibitem[{{Westmeier}(2018)}]{westmeier18}
{Westmeier} T (2018) {A new all-sky map of Galactic high-velocity clouds from
  the 21-cm HI4PI survey}. \mnras 474(1):289--299. \doi{10.1093/mnras/stx2757},
  {\href{https://arxiv.org/abs/1712.00909}{{arXiv:1712.00909}}} {[astro-ph.GA]}

\bibitem[{{Yang} et~al(2014){Yang}, {Hammer}, {Fouquet}, {Flores}, {Puech},
  {Pawlowski}, and {Kroupa}}]{yang14}
{Yang} Y, {Hammer} F, {Fouquet} S, et~al (2014) {Reproducing properties of MW
  dSphs as descendants of DM-free TDGs}. \mnras 442(3):2419--2433.
  \doi{10.1093/mnras/stu931},
  {\href{https://arxiv.org/abs/1405.2071}{{arXiv:1405.2071}}} {[astro-ph.GA]}

\bibitem[{{Yoshizawa} and {Noguchi}(2003)}]{yoshizawa03}
{Yoshizawa} AM, {Noguchi} M (2003) The dynamical evolution and star formation
  history of the small magellanic cloud: effects of interactions with the
  galaxy and the large magellanic cloud. \mnras 339(4):1135--1154.
  \doi{10.1046/j.1365-8711.2003.06263.x}

\bibitem[{{Zaritsky} et~al(2000){Zaritsky}, {Harris}, {Grebel}, and
  {Thompson}}]{zaritsky00}
{Zaritsky} D, {Harris} J, {Grebel} EK, et~al (2000) {The Morphologies of the
  Small Magellanic Cloud}. \apjl 534(1):L53--L56. \doi{10.1086/312649},
  {\href{https://arxiv.org/abs/astro-ph/0003155}{{arXiv:astro-ph/0003155}}}
  {[astro-ph]}

\bibitem[{{Zaritsky} et~al(2020){Zaritsky}, {Conroy}, {Naidu}, {Cargile},
  {Putman}, {Besla}, {Bonaca}, {Caldwell}, {Han}, {Johnson}, {Speagle}, and
  {Ting}}]{zaritsky20}
{Zaritsky} D, {Conroy} C, {Naidu} RP, et~al (2020) {Discovery of Magellanic
  Stellar Debris in the H3 Survey}. \apjl 905(1):L3.
  \doi{10.3847/2041-8213/abcb83},
  {\href{https://arxiv.org/abs/2011.09395}{{arXiv:2011.09395}}} {[astro-ph.GA]}

\bibitem[{{Zivick} et~al(2018){Zivick}, {Kallivayalil}, {van der Marel},
  {Besla}, {Linden}, {Koz{\l}owski}, {Fritz}, {Kochanek}, {Anderson}, {Sohn},
  {Geha}, and {Alcock}}]{zivick18}
{Zivick} P, {Kallivayalil} N, {van der Marel} RP, et~al (2018) {The Proper
  Motion Field of the Small Magellanic Cloud: Kinematic Evidence for Its Tidal
  Disruption}. \apj 864(1):55. \doi{10.3847/1538-4357/aad4b0},
  {\href{https://arxiv.org/abs/1804.04110}{{arXiv:1804.04110}}} {[astro-ph.GA]}

\bibitem[{{Zivick} et~al(2019){Zivick}, {Kallivayalil}, {Besla}, {Sohn}, {van
  der Marel}, {del Pino}, {Linden}, {Fritz}, and {Anderson}}]{zivick19}
{Zivick} P, {Kallivayalil} N, {Besla} G, et~al (2019) {The Proper-motion Field
  along the Magellanic Bridge: A New Probe of the LMC-SMC Interaction}. \apj
  874(1):78. \doi{10.3847/1538-4357/ab0554},
  {\href{https://arxiv.org/abs/1811.09318}{{arXiv:1811.09318}}} {[astro-ph.GA]}

\end{thebibliography}
}

\end{document}